\documentclass[twocolumn, letter]{aastex62}
\usepackage{multirow,amsfonts,amsmath,graphicx,graphics,natbib,subfigure,color,gensymb,longtable}

\newcommand{\ergcm}{erg cm$^{-2}$ s$^{-1}$}
\newcommand{\ergs}{erg s$^{-1}$}
\newcommand{\NH}{\mbox{$N_{\rm H}$}}
\newcommand{\NHgal}{\mbox{$N_{\rm H,MW}$}}

\newcommand\chandra{{\it Chandra}}

\newcommand{\xmm}{{\it XMM}}

\newcommand{\nustar}{\textit{NuSTAR}}

\graphicspath{{./}{figures/}}

\received{Month XX, 20XX}
\revised{Month XX, 20XX}
\accepted{Month XX, 20XX}
\submitjournal{ApJ}

\def \xmm {\textit{XMM-Newton}}
\def \at {AT\,2018cow}

\def\Msun{M$_{\odot}$}

\DeclareRobustCommand{\ion}[2]{\relax\ifmmode\ifx\testbx\f@series{\mathbf{#1\,\mathsc{#2}}}\else{\mathrm{#1\,\mathsc{#2}}}\fi\else\textup{#1\,{\mdseries\textsc{#2}}}\fi}
\shorttitle{AT\,2018cow: persistent X-rays?}
\shortauthors{Migliori et al.}

\begin{document}

%\title{ROARING TO SOFTLY WHISPERING: XMM-NEWTON OBSERVATIONS OF THE X-RAY DECAY OF AT\,2018cow}
\title{Roaring to softly whispering: Persistent X-ray emission at the location of the Fast Blue Optical Transient AT\,2018cow $\sim$3.7 yrs after  discovery and implications on accretion-powered  scenarios\footnote{Based on observations obtained with XMM-Newton, an ESA science mission with instruments and contributions directly funded by ESA Member States and NASA}}

\correspondingauthor{Giulia Migliori, Raffaella Margutti}
\email{giulia.migliori@inaf.it,rmargutti@berkeley.edu}

\author[0000-0002-0786-7307]{Giulia Migliori}
%\affil{Dipartimento di Fisica e Astronomia, Alma Mater Studiorum, Universit\`a degli Studi di Bologna, Via Gobetti 93/2, 40129 Bologna, Italy.}
\affiliation{INAF Istituto di Radioastronomia, via Gobetti 101, 40129 Bologna, Italy.}

\author[0000-0003-4768-7586]{R. Margutti}
\affiliation{Department of Astronomy, University of California, Berkeley, CA 94720-3411, USA}
\affiliation{Department of Physics, University of California, 366 Physics North MC 7300,
Berkeley, CA 94720, USA}

\author[0000-0002-4670-7509]{B. D. Metzger}
\affil{Department of Physics and Columbia Astrophysics Laboratory, Columbia University, New York, NY 10027, USA}
\affil{Center for Computational Astrophysics, Flatiron Institute, 162 5th Ave, New York, NY 10010, USA}

\author[0000-0002-7706-5668]{R. Chornock} 
\affiliation{Department of Astronomy, University of California, Berkeley, CA 94720-3411, USA}

\author[0000-0002-8853-9611]{C. Vignali}
\affiliation{Dipartimento di Fisica e Astronomia "Augusto Righi", Università di Bologna, via P. Gobetti 93/2 - 40129 Bologna, Italy}

\author[0000-0001-6415-0903]{D. Brethauer} %NuSTAR data cleaning
\affiliation{Department of Astronomy, University of California, Berkeley, CA 94720-3411, USA}

\author[0000-0001-5126-6237]{D.~L. Coppejans} 
\affiliation{Department of Physics, University of Warwick, Coventry CV4 7AL, UK} 

\author{T. Maccarone} 
\affiliation{Department of Physics \& Astronomy, Texas Tech University,
Box 41051, Lubbock, TX, 79409-1051, USA} 

\author[0000-0002-9396-7215]{L. Rivera Sandoval} 
\affiliation{Department of Physics and Astronomy, University of Texas Rio Grande Valley, Brownsville, TX 78520, USA}

\author[0000-0002-7735-5796]{J.~S. Bright}
\affiliation{Astrophysics, Department of Physics, University of Oxford, Keble Road, Oxford OX1 3RH, UK}
\affiliation{Department of Astronomy, University of California, Berkeley, CA 94720-3411, USA}

%\author[0000-0002-1984-2932]{B.~W. Grefenstette}
%\affiliation{Space Radiation Laboratory
%California Institute of Technology 
%1200 E California Blvd 
%Pasadena, CA 91125, USA}

%\author[0000-0000-0000-0000]{I. Vurm} 
%\affiliation{}

%\author[0000-0001-7081-0082]{M. Drout} 
%\affiliation{David A. Dunlap Department of Astronomy \& Astrophysics, University of Toronto}
%\affiliation{Dunlap Institute for Astronomy \& Astrophysics, University of Toronto}

%\author[0000-0003-0794-5982]{G. Terreran} 
%\affiliation{Las Cumbres Observatory, 6740 Cortona Drive, Suite 102, Goleta, CA 93117-5575, USA}
%\affiliation{Department of Physics, University of California, Santa Barbara, CA 93106-9530, USA}

%\author[0000-0002-8297-2473]{K.~D.~Alexander}
%\affiliation{Department of Astronomy/Steward Observatory, 933 North Cherry Avenue, Rm. N204, Tucson, AZ 85721-0065, USA}

\author[0000-0003-1792-2338]{T. Laskar}
\affiliation{Department of Physics \& Astronomy, University of Utah, Salt Lake City, UT 84112, USA}
\affiliation{Department of Astrophysics/IMAPP, Radboud University, P.O. Box 9010, 6500 GL, Nijmegen, The Netherlands}

\author[0000-0002-0763-3885]{D. Milisavljevic} 
\affiliation{Purdue University, Department of Physics and Astronomy}
\affiliation{Integrative Data Science Initiative, Purdue University}

\author[0000-0002-9392-9681]{E. Berger} 
\affiliation{Center for Astrophysics, Harvard \& Smithsonian}

\author[0000-0002-8070-5400]{A.~J. Nayana} 
\affiliation{Department of Astronomy, University of California, Berkeley, CA 94720-3411, USA}
\affiliation{Indian Institute of Astrophysics, II Block, Koramangala, Bangalore 560034, India}

%\author[0000-0000-0000-0000]{C. Guidorzi} 
%\affiliation{}

%\author[0000-0000-0000-0000]{P. Blanchard} 
%\affiliation{}

%\author[0000-0002-2555-3192]{M. Nicholl}
%\affiliation{Astrophysics Research Centre, School of Mathematics and Physics, Queens University Belfast, Belfast BT7 1NN, UK}

%\author[0000-0002-9392-9681]{E. Berger} 
%\affiliation{Center for Astrophysics, Harvard \& Smithsonian}

%\author[0000-0000-0000-0000]{A. Hajela} 
%\affiliation{}

%\author[0000-0000-0000-0000]{A.~J. Nayana} 
%\affiliation{}

%CO-AUTHORS of the XMM proposal  (they have yet to contribute to this paper, I ssuggest alphabetical order) ---
%Brian Grefenstette bwgref@srl.caltech.edu
%Indrek Vurm indrek.vurm@gmail.com
%Maria Drout  maria.drout@utoronto.ca 
%Dan Milisavljevic dmilisav@purdue.edu
%Giacomo Terreran gterreran@lco.global
%Kate Alexander  kdalexander@arizona.edu
%Tanmoy Laskar tanmoylaskar@gmail.com
%Cristiano Guidorzi guidorzi@fe.infn.it
%Peter Blanchard peter.blanchard@northwestern.edu
%Matt Nicholl  matt.nicholl@qub.ac.uk 
%Edo Berger eberger@cfa.harvard.edu
%Aprajita Hajela aprajita.hajela@gmail.com
%To add: Nayana AJ nayana@berkeley.edu Joe Bright

%\author{}
%\affiliation{}

\begin{abstract}
We present the first deep X-ray observations of a luminous FBOT \at{}\, at $\sim 3.7\,\rm{yr}$ since discovery, together with the re-analysis of the observation at $\delta t \sim 220$\,d. 
 X-ray emission is significantly detected at a location consistent with \at{}.
The very soft X-ray spectrum and sustained luminosity are  distinct from the spectral and temporal behavior of the LFBOT in the first $\sim100$ days, and would possibly signal the emergence of a new emission component, although a robust association with \at{} can only be claimed at $\delta t \sim 220$\,d, while at $\delta t \sim 1350$\,d contamination of the host galaxy cannot be excluded. We interpret these findings in the context of the late-time panchromatic emission from \at{}, which includes the detection of persistent, slowly-fading UV emission with $\nu L_{\nu}\approx 10^{39}\,\rm{erg\,s^{-1}}$. Similar to previous works, (and in analogy with arguments for Ultra-Luminous X-ray sources --ULXs), these late-time observations are consistent with thin-disks around Intermediate Mass Black Holes (IMBHs,  with $M_{\bullet}\approx 10^3-10^4\,\rm{M_{\sun}}$) accreting at sub-Eddington rates. However, differently from previous studies, we find that smaller-mass BHs with $M_{\bullet}\approx 10-100\,\rm{M_{\sun}}$ accreting at $\gtrsim$ the Eddington rate cannot be ruled out, and provide a natural explanation for the inferred compact size ($R_{\rm out}\approx 40\,R_{\sun}$) of the accretion disk years after the optical flare. Most importantly, irrespective of the accretor mass, our study lends support to the hypothesis that LFBOTs are accretion-powered phenomena and that, specifically, LFBOTs constitute electromagnetic manifestations of super-Eddington accreting systems that evolve to $\lesssim$ Eddington over a $\approx 100$\,days time scale. 

\end{abstract}


%%%%%%%%%%%%%%%%%%%%%%%%%%%%%%%%%%%%%%%%%%%%%%%%%%%%%%%%%%%%
\section{Introduction} \label{sec:intro}
High-cadence, wide-area optical surveys have recently led to the discovery of a new class of luminous transients (peak bolometric luminosity  $L_{\rm pk}\approx10^{41}-10^{44}$ \ergs), characterized by rapid rising times ($t_{\rm rise}<10$\,d) and  blue colors that are signatures of high effective temperatures $>10^4$\,K. The class of Fast Blue Optical Transients \citep[FBOTs,][]{Drout+2014}\footnote{Also referred to in the literature as Fast Evolving Luminous Transients \citep[FELTs,][]{Rest18}.} collects a few tens of systems \citep[e.g.,][]{Poznanski10,Drout+2014,Arcavi16,Pursiainen18,Tampo+2020,Ho22sample}. While the observational properties of FBOTs are heterogeneous and most likely reflect some intrinsic diversity among the transients, their fast rise times, high luminosities and lack of ultraviolet (UV) line blanketing cannot be easily explained with the radioactive decay of newly-synthesized $^{56}$Ni as for ordinary supernova (SN) explosions. This argument especially applies to the most optically luminous end of the FBOT population (Luminous FBOTs --LFBOTs -- hereafter), which shows a combination of extreme peak luminosities reaching $L_{\rm pk}\approx10^{44}$ \ergs combined with the shortest rise-times of just a few days. 

The key observational properties and inferences on the sub-class of LFBOTs are summarized below.
(i) Differently from FBOTs, LFBOTs are intrinsically rare events (local volumetric rate  $\lesssim 1\%$ of the core-collapse SN rate; \citealt{Coppejans+20,Ho20AT2018lug}). (ii) These studies also revealed that  LFBOTs have luminous X-ray and radio counterparts, sometimes as luminous as  Long Gamma-Ray Bursts (LGRBs), and that,  similar to GRBs, (iii) LFBOTs are capable of launching relativistic outflows, which implies the presence of a compact object (BH or NS, pre-existing or newly formed). (iv) LFBOTs show highly time-variable non-thermal X-ray (and optical) emission \citep{Riv18,Margutti+19,Ho22opticalVar}, similar to GRB afterglows and TDEs, and clearly distinct from X-ray SNe that are powered by the shock interaction with the circumstellar medium (CSM). (iv) LFBOTs are surrounded by dense but radially confined CSM, as revealed by their rapidly declining radio light-curves \citep{Bright+22, Nayana21cow,Ho+20,Ho20AT2018lug} and  potential infrared ``dust echo'' signatures \citep{Metzger23}. (v) In stark contrast with GRBs, the optical spectra of LFBOTs show the presence of H, demonstrating that LFBOTs are H-depleted but not H-free (e.g., \citealt{Perley+19,Margutti+19}). (vi) Finally, LFBOTs preferentially occur in low-mass star-forming  galaxies, thus showing a preference for low-metallicity environments, and hence suggesting a connection with massive stars \citep{Coppejans+20,Lyman+20, Yao21}.\footnote{The lack of evidence of cooling of LFBOTs like AT2018cow makes them phenomenologically distinct from the new class of luminous, fast-cooling transients, which occur in early-type galaxies and have no detected X-ray or radio emission \citep{Nicholl23}. }

Physical scenarios to explain the LFBOT population fall under two broad categories: those invoking the interaction of a shock with a dense CSM as a way to efficiently convert the outflow's kinetic energy into radiation; and those involving the presence of energy injection by a ``central engine'' (e.g., \citealt{Prentice+2018,Margutti+19,Perley+19}).
In this second class of models, LFBOTs could be powered by  a newly-formed rapidly-rotating magnetar (e.g., \citealt{Vurm&Metzger21}), an accreting black hole (BH; e.g., \citealt{AS2021,Gottlieb+2022}) born in a failed blue supergiant star explosion \citep{Quataert+19,Antoni22}, as well as tidal disruption events (TDEs) by intermediate-mass black holes (IMBHs), e.g., \cite{Kuin+19,Perley+19}. Recently, \citet{Metzger22} showed that LFBOTs could also result from the binary merger of a Wolf-Rayet star with its BH or NS companion. While pure CSM interaction models (e.g., \citealt{Fox&Smith19,Schroder+20,Dessart+21interaction,Leung+21,Margalit22,Pellegrino+2022}) struggle explaining the mildly-relativistic ejecta, rapid variability time scales and non-thermal X-ray spectra of LFBOTs, it is important to note that the two sets of models (i.e., ``interaction'' vs. ``central engine'') are not mutually exclusive, and different parts of the spectrum can be dominated by different physical mechanisms (e.g., the radio emission could be shock-CSM interaction powered, while the rest of the spectrum is not).

Located in the spiral arm of the dwarf star-forming galaxy CGCG 137$-$068 at a distance $d\sim 60$\,Mpc, \at{}\, is the nearest LFBOT  discovered so far (\citealt{Prentice+2018,Smartt+2018a,Perley+19}) and offers an unparalleled opportunity to test and constrain the scenarios above. \at{}\, has been extensively observed across the entire electromagnetic spectrum \citep{Riv18,Kuin18,Perley+19,Ho+19,Margutti+19, Nayana21cow}. 
The observational findings from these campaigns that are most relevant to our study are summarized below.

\at{} displayed luminous, highly-variable X-ray emission (peak $L_x>10^{43}$ \ergs) of \emph{non-thermal} origin with two spectral components: a persistent, relatively hard power-law ($F_{\nu}\propto \nu^{-0.6}$) at $h\nu>$0.1 keV, and a transient Compton-hump feature dominating  at $h\nu>$10 keV at $\delta t<10$ d \citep{Margutti+19,Riv18}. While the bright radio-to-mm emission of \at{}\, stems from a non-relativistic shock interaction with a dense medium \citep{Margutti+19,Ho+19,Nayana21cow}, the persistent optical-UV blue colors requiring $T_{\rm{eff}}\approx15000$\,K weeks after discovery (e.g., \citealt{Margutti+19,Perley+19,Xiang21-18cow}), and the broad-band X-ray properties above %\footnote{See however \cite{Riv18}. We also note that in \cite{Fox19}, the X-ray spectrum of \at{} is \emph{assumed} to be of thermal origin, but no quantitative spectral modeling is attempted. } 
demand a different powering source, which can be in the form of a long-lived central engine. Slowly-decaying, luminous ($L>10^7\,\rm{L_{\sun}}$) UV emission years after discovery has been recently reported by \cite{Sun22,Sun22b,Chen2023b,Chen2023a,Mummery23,Inkenhaag23}, and similarities have been noted with radiation powered by accretion processes on compact objects. Along the same lines, a high-frequency (224 Hz) quasi-periodic oscillation (QPO) feature in the X-ray timing properties of \at{}\,
%in the average power density spectrum of \at{} over the 60-day outburst 
supports the presence of a compact object, either a NS or BH with mass $M<$850\,\Msun \citep[][but see also \citealt{Zhang+2022}]{DJ21}. While no evidence was found for a long-lived relativistic jet such as those of GRBs \citep{Bietenholz+2020,Mohan+20}, panchromatic observations of \at{}\, further indicated a complex geometry that strongly departs from spherical symmetry, as was directly confirmed by the very large optical polarization ($\sim7\%$) at early times   \citep{Maund23}.

Here we present deep X-ray observations of \at{}\ performed up to $\approx3.7$ yrs after the optical discovery, and we discuss the implications of a late-time X-ray detection on the intrinsic nature of this new class of transients. 
The paper is organized as follows: in \S\ref{sec:XMM} and \S\ref{sec:NuSTAR} we present the analysis of the \emph{XMM-Newton} and {\it NuSTAR} observations at $\delta t\geq$218\,d. We put the X-ray data into the broader context of the late-time emission from \at{}\, in \S\ref{sec:nature} and \S\ref{sec:SED}, and we discuss our findings in \S\ref{sec:disc2}. We conclude in \S\ref{Sec:SummaryConclusions}. We provide in Appendix \ref{sec:Chandra} the analysis of {\it Chandra} HETG observations at $\delta t=8.2$\,d that has not been published elsewhere. Times are referred to the epoch of the optical discovery, which is MJD 58285.44, and we adopt a luminosity distance of 60 Mpc \citep{Prentice+2018,Smartt+2018a,Perley+19}.

%%%%%%%%%%%%%%%%%%%%%%%%%%%%%%%%%%%%%%%%%%%%%%%%%%%%%%%%%%%%%%%%%%%%%%
\section{Late-time XMM-Newton follow-up} \label{sec:XMM}

\begin{figure*}
\centering
 \includegraphics[trim=1.3cm 10.5cm 1.1cm 10cm, clip=true]{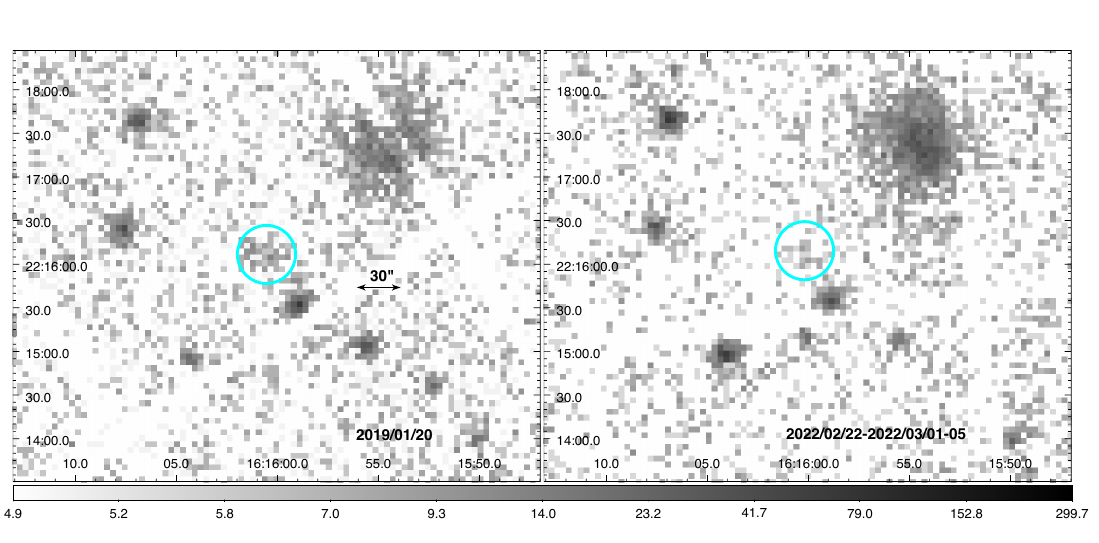}
 \caption{Mosaic of the 0.2--4.5 keV images of \at~in the three EPIC cameras for Obs3 (\emph{left panel}) and Obs4 (\emph{right panel}). Color intensity scales logarithmically with the number
of counts. The cyan circle ($r=20\arcsec$) marks the \xmm{} detection.}
 \label{fig:f1}
 \end{figure*}

Since its discovery, \at~was targeted four times by  \xmm\, (PI Margutti). The first two epochs (Obs1 and Obs2) were presented in \citet{Margutti+19}. Here we report the analysis of the 2019 observation (Obs3, $\delta t\sim$218 d, Obs ID 0822580601) and the 2022 one (split in three, $\delta t\sim$1349 d to 1358 d; Obs4a, Obs4b, Obs4c, collectively Obs4; Obs IDs: 0843550401, 0843550501, 0843550601), that sample the very late time X-ray evolution of an FBOT  (Table \ref{tab:log}).

We reduced and analyzed the data of the three European Photon Imaging Camera (EPIC)-pn, MOS1 and MOS2 using the Scientific Analysis System (SAS) v. 20.0.0 and calibration files CALDB 3.13.
Obs4b and Obs4c were significantly impacted by high-background flares while Obs3 and Obs4a were not. For the former, optimized flare filtering was achieved with  {\tt\string espfilt\footnote{\url{https://xmm-tools.cosmos.esa.int/external/sas/current/doc/espfilt/espfilt.html}}}, which led to a severe reduction in the effective exposure time, especially for the pn camera (Table \ref{tab:log}). 

\subsection{Source Detection}
 An X-ray source is clearly visible at the location of \at{} in the 0.2--10 keV image of Obs3 (see Figure \ref{fig:f1}). 
We restricted our analysis to the 0.2--4.5 keV energy range (i.e., band 1 to 4 of the \emph{XMM} catalogues) since above $\approx 5$\,keV the background dominates. We estimated the background contribution from several source-free circular regions
%For the background, we selected multiple, free-from-source, circular regions 
around \at~and in the same chip. In  the 0.2-4.5 keV pn image, which has the best statistics, we measured 41$\pm$14 net counts in a 20\arcsec~ circular region centered at the optical coordinates of \at.  To assess the significance of the detection   we ran the task {\tt\string edetect\_chain} for all the three cameras simultaneously over the full 0.2--4.5 keV energy band, and in four sub-energy bands (i.e., 0.2--0.5 keV, 0.5--1.0 keV, 1.0--2.0 keV and 2.0--4.5 keV). The source is significantly detected in the full 0.2--4.5 keV with a resulting detection maximum likelihood  {\tt\string DET\_ML=}10.8 ($\gtrsim 3\,\sigma$, Gaussian equivalent). The 0.2--4.5 keV net count-rates for each instrument are reported in Table \ref{tab:log}.
The analysis of the distribution of the counts in the energy sub-bands shows that most counts are clustered below 2 keV.

The centroid coordinates of the X-ray source in Obs3 are RA:244.00104 deg, DEC: 22.26758 deg with a positional uncertainty of 1.5\arcsec, fully consistent with the coordinates of the optical counterpart. Note that the X-ray source closest to \at~is at 38\arcsec~and its flux did not vary through the \xmm~observations, suggesting a negligible level of contamination.
The detection of \at~in \emph{XMM} Obs3 has  been previously reported by \citet{Yao21}\footnote{This work focuses on the FBOT AT2020mrf, the detection of \at{} is reported in Appendix A.}. A source is present also in 
%Moreover, a detection for this observation is also present in 
the fourth \emph{XMM-Newton} serendipitous source catalog \citep[4XMM, data release 11, DR11][]{webb20}, instead the non-detection claimed in \citet{DJ21} is likely due to the use of the MOS1 image alone.

From visual inspection, no source is apparent  in the individual images of the three  exposures of Obs4. %In order to search for our source in all Obs4 pointings, 
To maximize our sensitivity to faint sources we ran the SAS task  {\tt\string edetect\_stack}, which %allows to 
performs source detection on stacked images from different exposures
%stacked source detection on multiple observations  
\citep[][]{Trau19,Trau20}. Following a similar argument as for Obs3, we restricted our analysis to $\le 5$ keV.
We find evidence for a point-like\footnote{The task assigns a zero source extent if the likelihood of the
source being extended falls below a threshold of four or its extent
radius below 6\arcsec \citep{Trau19}.} X-ray source %is present 
with a combined EPIC detection likelihood\footnote{This is defined as equivalent likelihood, i.e. the detection likelihood of the  source in $n$ individual images, see Eq.3 in \cite{Trau19}.} {\tt\string DET\_ML=}9.1 ($3.8\,\sigma$ Gaussian equivalent) in the 0.2--4.5 keV energy band. However, the source position is displaced  4.4\arcsec~north (RA:244.00079 deg, DEC: 22.26925 deg) with respect to the optical position of \at{}. The 1-$\sigma$ positional error is 1.9\arcsec, and \citet{Trau19} quote a mean systematic error in the range 0.43\arcsec-0.73\arcsec,  hence an offset  of $\sim$2$\sigma$ from the optical position. We note that the X-ray position is also 5.5\arcsec\,  offset from the host galaxy centroid. Inspection of the {\tt\string edetect\_stack} intermediate steps shows that this offset is the result of the final maximum likelihood fitting performed by the {\tt\string eml\_detect} task, while the input position provided in the previous step by the {\tt\string ebox\_detect} task is within 2\arcsec~from the \at~position, suggesting that uncertainties in merging the images, or performing the PSF fitting may affect the final position \citep[see e.g.][]{Rosen2016,webb20}. %We further discuss this point in \ref{sec:XX}. \textcolor{red}{XX}
The 0.2--4.5 keV net count-rates are reported in Table \ref{tab:log}.

\subsection{X-ray Spectral Analysis}\label{sec:Xrayfit}
For Obs3, we extracted source spectra using data acquired by the three cameras. We used a source region with radius defined as the best-fitting radius for each camera determined by the detection tool and selected multiple, contiguous and source-free regions for the background. We employed the W-stat statistic and simultaneously fit the source and background spectra. 
We tested two spectral models: a power-law and a black-body model, both combined with an absorption component (i.e., \texttt{tbabs*pow} and \texttt{tbabs*bbody}). Unsurprisingly, given the low-number counts, the two models are statistically indistinguishable. We derived a very soft best-fitting power-law photon index of  $\Gamma=2.9^{+0.6}_{-0.4}$, whereas we obtained a best-fitting black-body temperature of $kT=0.16^{+0.04}_{-0.03}$ keV. In both cases, no evidence for intrinsic absorption is found and we thus froze  the \NH\, to the  
Galactic value \NHgal=0.05$\times 10^{22}\,\rm{cm^{-2}}$ \citep{Kalberla05}. This is in line with upper limits derived from high-count spectra at early times \citep[ \NH$_{\rm ,int}<$0.02$\times 10^{22}\,\rm{cm^{-2}}$,][]{Margutti+19}. We further explored a multi-color black-body disk model (\texttt{diskbb}), which is the implementation of the standard \cite{Shakura&Sunyaev73} thin-disk model. The inner disk  temperature is constrained to $T_{in}=$0.23$^{+0.06}_{-0.04}$ keV, while for the inner disk radius we infer a maximum value, $R_{in}\lesssim $1340 km for a face-on disk. The UV flux predicted by this model fails to explain the observed emission from \at\, and we refer the reader to \S\ref{sec:SED} for a self-consistent UV-to-X-ray multi-color black-body disk modeling.
Fitting with a thermal plasma model (\texttt{apec}) leads to an extremely low temperature ($kT\sim$0.009 keV) and to a worse fit.

To conclude, while the Obs3 data cannot constrain the spectral shape in detail, there is a clear indication of a spectral softening of the X-ray emission at $\delta t \ge 210$\,d, which signals a change with respect to the persistently hard X-ray emission with $\Gamma\sim1.5$ observed at $\delta t\le 82$\,d (\citealt{Riv18,Margutti+19}, Fig. \ref{fig:f2}). The 0.3--10 keV unabsorbed flux inferred from the spectral analysis ranges from (3.1$^{+3.5}_{-1.6})\times$10$^{-15}$  \ergcm\,  (black-body model) to (4.5$\pm$1)$\times$10$^{-15}$  \ergcm\, (power-law model). Note that the higher flux of $F_x\sim$1.6$\times$10$^{-14}$ \ergcm\, reported in \citet[][]{Yao21} for this observation is the result of the harder $\Gamma\approx$2.0 that was \emph{assumed}  by the authors.

While the limited statistics of Obs4 also leave the spectral models unconstrained, we note that most of the counts in the source region are at energies $\le$2 keV, which supports the conclusion of spectral softening of the source. Using the output of the detection algorithm, we estimated the hardness ratios (HR) in the 0.2-1 keV (b1), 1-2 keV (b2) and 2-4.5 keV (b3) energy bands for the pn data. For $HR1=(b2-b1)/(b2+b1)$ and $HR2=(b3-b2)/(b2+b3)$, we obtained $HR1=-0.4\pm0.5$ and $HR2=-0.1\pm0.5$, respectively, which are in rough agreement with the values in Obs3 ($HR1=-0.9\pm0.4$ and $HR2=-1.0\pm0.6$), although, given the large uncertainties, the HR measurements should be considered as purely indicative.
Assuming $\Gamma=2.9$ as derived from Obs3, the pn count rate of Obs4 converts into an unabsorbed 0.3--10 keV flux of $F_x\approx$(1.0$\pm$ 0.4)$\times$10$^{-15}$ \ergcm\, (Table \ref{tab:log}).

%%%%%%%%%%%%%%%%%%%%%%%%%%%%%%%%%%%%%%%%%%%%%%%%%%%%%%%%%%%%%
\section{Late-time NuSTAR follow-up}
\label{sec:NuSTAR}
We obtained deep observations of \at{}\, with the \emph{Nuclear Spectroscopic Telescope Array} \citep{Harrison13} on 2022 March 2
%this is MJD= 59640, so 59640-58285.44= 1354.6 days since discovery
(Program \#084355, PI Margutti, $\delta t=1354.6$\,d ). \emph{NuSTAR} observations were processed using \texttt{NuSTARDAS} v1.9.7 and the \emph{NuSTAR} \texttt{CALDB} released on 2022 May 10. Part of our observations were severely affected by Solar activity. Filtering out periods of increased detector background with \texttt{saacalc=3, saamode=strict tentacle=yes} leads to effective exposures of $\approx 113$ ks and $\approx 111$ ks on module A and B, respectively.  No source of significant hard X-ray emission is detected at the location of \at{}. Using  extraction regions that sample $50\%$  of the \emph{NuSTAR} Point Spread Function (PSF) and centered at the optical position of \at{}, we infer a combined count-rate upper limit of  $1.1\times 10^{-4}\,\rm{c\,s^{-1}}$ (10--79 keV). This translates into a flux limit of $F_{x,hard}<1.3\times 10^{-14}$ \ergcm\,($L_{x,hard}<5.\times 10^{39}\,\rm{erg\,s^{-1}}$) for an assumed spectral power-law index of $\Gamma=2.9$ that best fits the \xmm\, data (Obs3, see \S\ref{sec:XMM}). 

%%%%%%%%%%%%%%%%%%%%%%%%%%%%%%%%%%%%%%%%%%%%%%%%%%%%%%%%%%%%%%%%%%%%%%%%%%%%%%%%%%%%%%%
\section{Nature of the late-time X-ray emission}
\label{sec:nature}

\subsection{Transient Emission vs. Star Formation } 
Before proceeding further, we discuss the origin of the late-time X-ray emission, whether it can be ascribed to \at{} and in what fraction.
In non-active galaxies, like the host of \at{}, a possible source of contamination is represented by the X-ray emission of (i) X-ray binaries (XRBs), both low-mass and high-mass XRBs (LMXRBs and HMXRBs, respectively), which are dominant in the 2--10 keV energy range; and (ii) the hot ISM, mainly relevant below 2 keV. The X-ray luminosities of these components correlate with the star formation rate \citep[SFR, see e.g.][and references therein]{Mineo12,Mineo2014,Lehmer2016}. We adopted the empirical $L_X$-SFR relations in \citet{Lehmer2016}, which scale with redshift,  SFR and stellar mass ($M_\ast$), and used the SFR and $M_\ast$ inferred for the host galaxy of \at{}\, from the SED fitting of the optical data in \citet{Perley+19}: SFR$=$0.22$^{+0.03}_{-0.04}$ \Msun{} yr$^{-1}$ and $M_{\ast}=$(1.42$^{+0.17}_{-0.25})\times 10^9$ \Msun{} (also in agreement with \citealt{Michalowski19} and \citealt{Lyman+20}). 
We estimated a luminosity range $\sim 6-9\times 10^{38}$ \ergs\, for each, 0.5--2 keV and 2--10 keV, band \citep[similar values are obtained using the relation in][]{Mineo2014}. The estimated 1$\sigma$ scatter in the relations is 0.17 dex, although the galaxy-to-galaxy spread in the 2-10 keV luminosity could be larger, up to 0.4 dex, and sensitive to variations in metallicity, stellar age and XRB populations \citep{Lehmer2016}.

The level of the observed emission is thus of the same order of the SF estimates. However, we note that at $\delta t=$218 d (Obs3): (i) the emission is significantly brighter than Obs4, thus showing the limitations of the SFR-based inferences; moreover (ii) the centroid of the X-ray emission is consistent with the position of \at{} (and 6.8\arcsec\,from the host galaxy core); (iii) the typical star forming spectral components \citep[a thermal model with $kT\sim$0.3--2 keV and a power law with $\Gamma\sim$1.8--2.0, see e.g.][]{Lehmer2016} cannot satisfactorily model the 
 the X-ray spectrum.
For Obs4 ($\delta t=$1349$-$1358 d), the poor characterization of the X-ray emission in terms of localization and spectral properties makes it difficult to be conclusive between \at{} and a star formation origin. 
As a further attempt, we used the \chandra{} observation taken at $\delta t=$8.2 d (see Appendix A) and calculated the X-ray flux in the host galaxy region (excluding the \at{} emission). Unfortunately, the 90\% c.l. upper limit on the 0.3--10 keV unabsorbed flux is 2.4$\times$10$^{-13}$ \ergcm, corresponding to a 0.3--10 keV luminosity of $\leq$10$^{41}$ \ergs, not sufficiently deep to probe the expected SF luminosities.

In the following we consider two possible scenarios: (i) either the X-ray emission at $\delta t$=1349--1358 d {\bf (Obs4)} is still mostly contributed by \at{}, or (ii) \at{} had faded below the detectable level and we are detecting the SF-related X-ray emission from the host galaxy. In both cases, the hard X-ray emission from \at{}\, at this epoch is $L_{x,hard}<6\times 10^{39}\,\rm{erg\,s^{-1}}$(10--79 keV).
 Furthermore, if (ii), the SF component could contribute up to $\approx 25\%$ of the best-fit X-ray flux measured at $\delta t=$218 d {\bf (Obs3)}. This raises the question whether the SF contamination could be responsible of the  spectral softening observed in Obs3. Incidentally, we note that a typical SF X-ray spectrum is not expected to be as soft as the observed one \citep[see e.g. the template presented in Figure 4 of ][]{Lehmer2016}. As a further test, we simulated an X-ray spectrum and supposed the (pessimistic) scenario of a total flux, at the time of Obs3, of $2.8\times10^{-15}\,\rm{erg\,cm^{-2}\,s^{-1}}$ and a true host flux of $1.4\times10^{-15}\,\rm{erg\,cm^{-2}\,s^{-1}}$ (both values within $\sim$1$\sigma$ of the measured/estimated values), meaning that fully half of the flux detected in Obs3 would be from the host galaxy. For the host galaxy spectrum, we referred to the mean X-ray SED of local star-forming galaxies presented in \citet{Lehmer2016}: a thermal model with $kT=0.5$ keV was assumed for the hot gas and a power-law with $\Gamma=1.9$ for the XRB emission, while the relative contributions were set so that the former(/latter) dominates below(/above) $\sim$1.5 keV \citep[see Figure 4 in ][]{Lehmer2016}. The X-ray spectrum of \at{} was modelled with a power-law with a photon index equal to 1.5, i.e. the value measured at early times. The underlying assumption is that there is no spectral evolution in the X-ray emission of \at. Fitting the simulated X-ray spectrum with a composite model that accounts for all contributions, we obtained for \at{} a best-fit photon index $\Gamma=1.5\pm0.7$, while a fit with a simple power-law model corrected for Galactic absorption, as done for Obs3 spectrum, gives a best-fit photon index $\Gamma=1.8\pm 0.3$. This is below the observed range measured in Obs3, albeit the large uncertainties, thus supporting the indications of a true spectral softening of the X-ray emission of \at.

%%%%%%%%%%%%%%%%%%%%%%%%%%%%%%%%%%%%%%%%%%%%%%%%%%%%%%%%%%%%%%%%%%%%%%%%%%%%%%%%%%%%%%%

\subsection{\at~X-ray Light-Curve }
\begin{figure*}
\centering
\includegraphics[trim=0.cm 0.cm 0.cm 0.cm, clip=true,angle=-90,width=1.6\columnwidth]{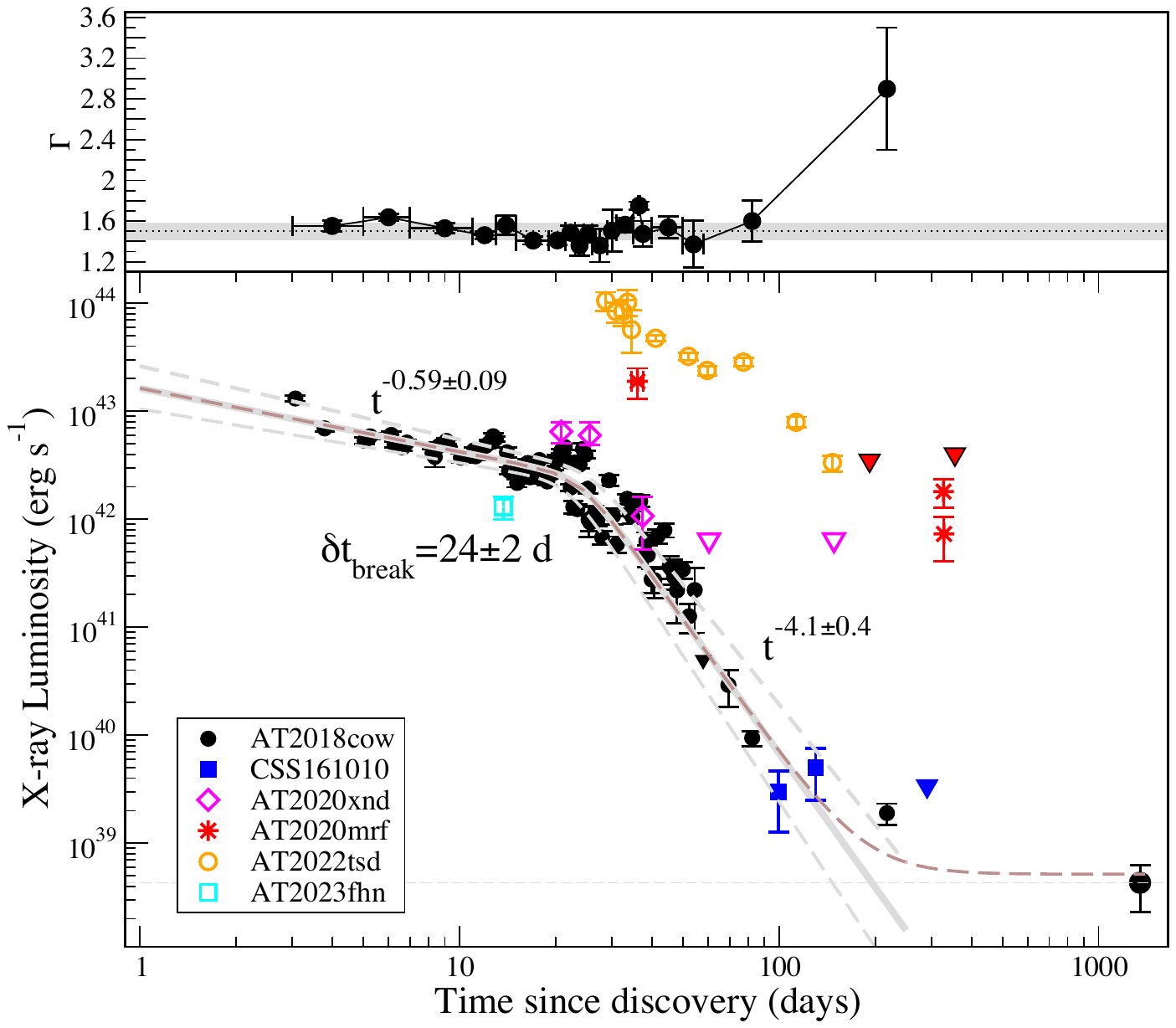}\hspace{7mm}
\caption{ Temporal evolution of \at~at X-ray energies (0.3-10 keV), as captured by Swift-XRT, XMM-Newton and Chandra. X-rays are modeled with a power-law spectrum. In the \emph{upper panel} the measured X-ray photon indexes are reported: $\Gamma\sim$1.5 at early time, while the 218 d spectrum suggests a strong softening, $\Gamma\sim$2.9. No measurement of $\Gamma$ was possible for the last XMM observation. The dotted line and grey shaded area are the averaged photon index at $\delta t\lesssim$83 d and the relative uncertainty. In the \emph{lower panel}, the gray solid line is the best-fit broken power-law decay assuming that \at{} has faded after $\delta t\lesssim$218 d, the short-dashed grey lines being the uncertainties on the model parameters. The long-dashed brown line is the broken power-law with the addition of a constant component, mimicking a late-time flattening of the lightcurve, under the hypothesis that the X-ray flux at $\delta t=$1349$-$1358 d is associated with \at{} (see text). The five X-ray detected FBOTs, CSS161010 \citep{Coppejans+20}, AT2020xnd \citep{Bright+22,Ho+22}, AT2020mrf \citep{Yao21}, AT2022tsd \citep{Matthews23} and AT2023fhn \citep{Chrimes23} are also shown for comparison. }
\label{fig:f2}
\end{figure*}

The temporal evolution of the soft X-ray flux at the location of \at{}\, is remarkable: as shown in Figure \ref{fig:f2}, at $\delta t=$218 d the 0.3--10 keV $L_x$ has dropped by $\approx$4 orders of magnitude with respect to the time of the discovery (from $\sim$10$^{43}$ \ergs\ to $\sim2\times$10$^{39}$ \ergs). However, between $\delta t=$218 d and $\delta t=$1349$-$1358 d the X-ray flux decreases by a factor $\lesssim$4. We fit the  X-ray lightcurve at $\delta t\le$218\,d with a smoothed broken power-law model, thus assuming that beyond that epoch we are detecting X-ray emission from the host galaxy. The best fit model (Figure \ref{fig:f2}) indicates an initial decay $L_x\propto\,$t$^{-0.59\pm0.09}$, which drastically steepens to $L_x\propto\,$t$^{-4.1\pm0.04}$ at $\delta t_{break}\sim$24$\pm$2 days. Note that this model has the goal to reproduce the long-term evolution of the X-ray emission, not the superposed hour-to-day temporal variability, evident in from the residuals, which was discussed in previous works \citep{Riv18,Margutti+19}. As the spectrum softens, the observed $L_x$ at $\delta t= 218$\,d appears in excess ($\sim$2.7$\sigma$) with respect to the broken power-law model.  In the hypothesis that the emission is instead still associated with \at{}, a new component, here chosen to be constant with $L_x= 4\times 10^{38}\,\rm{erg\,s^{-1}}$, needs to be added to the X-ray lightcurve model (smoothed broken power-law$+$constant) to match the last \xmm\, epoch.
We will discuss the implications of a persistent vs. rapidly fading X-ray emission in FBOTs at $\delta t\gtrsim 100$\,d in \S\ref{sec:disc2}.

\begin{figure*}
\centering
\includegraphics[trim=0.5cm 1.5cm 6.5cm 2.5cm, clip=true,width=1.\columnwidth]
{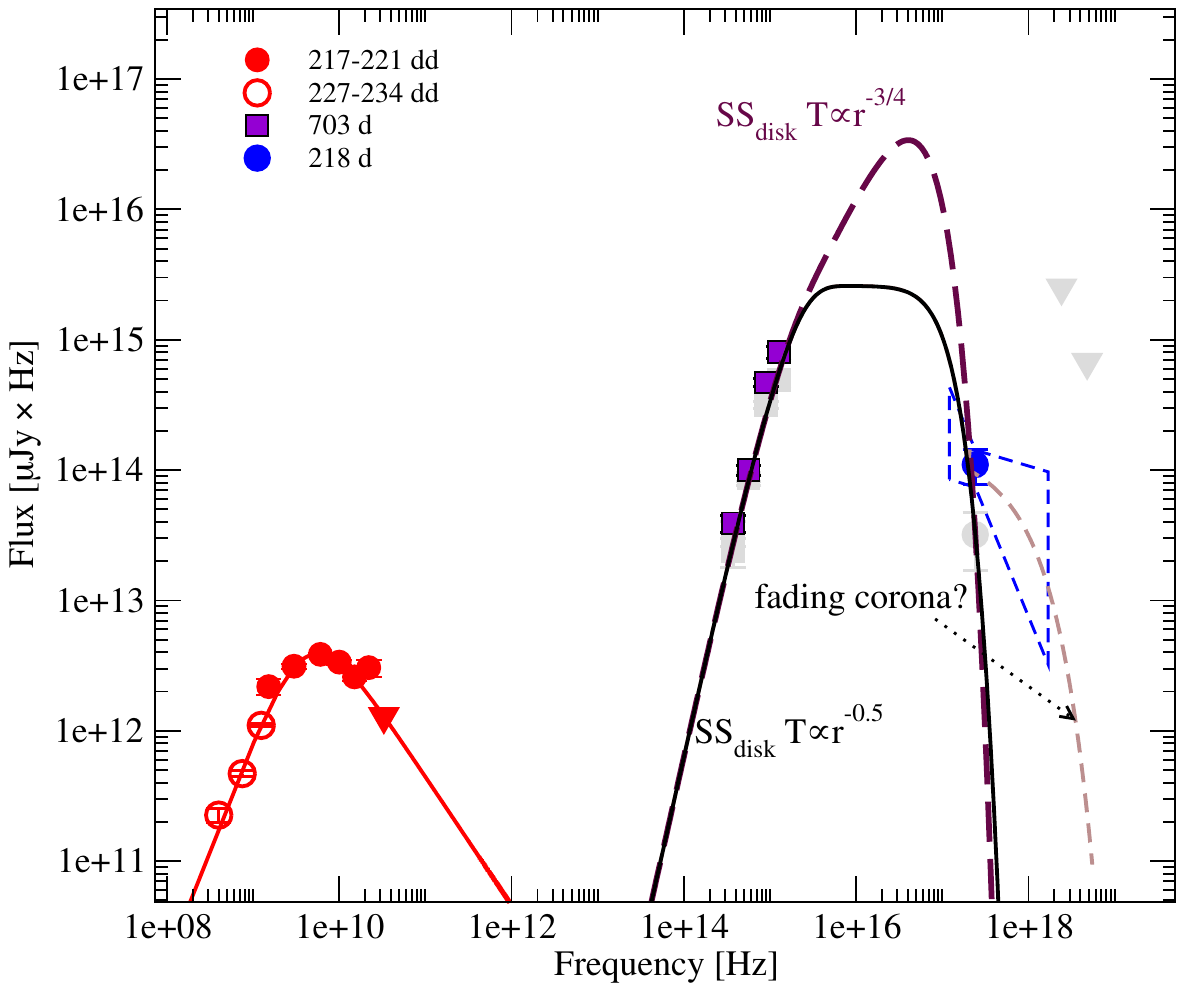}
\includegraphics[trim=0.5cm 1.5cm 6.5cm 2.5cm, clip=true,width=1.\columnwidth]{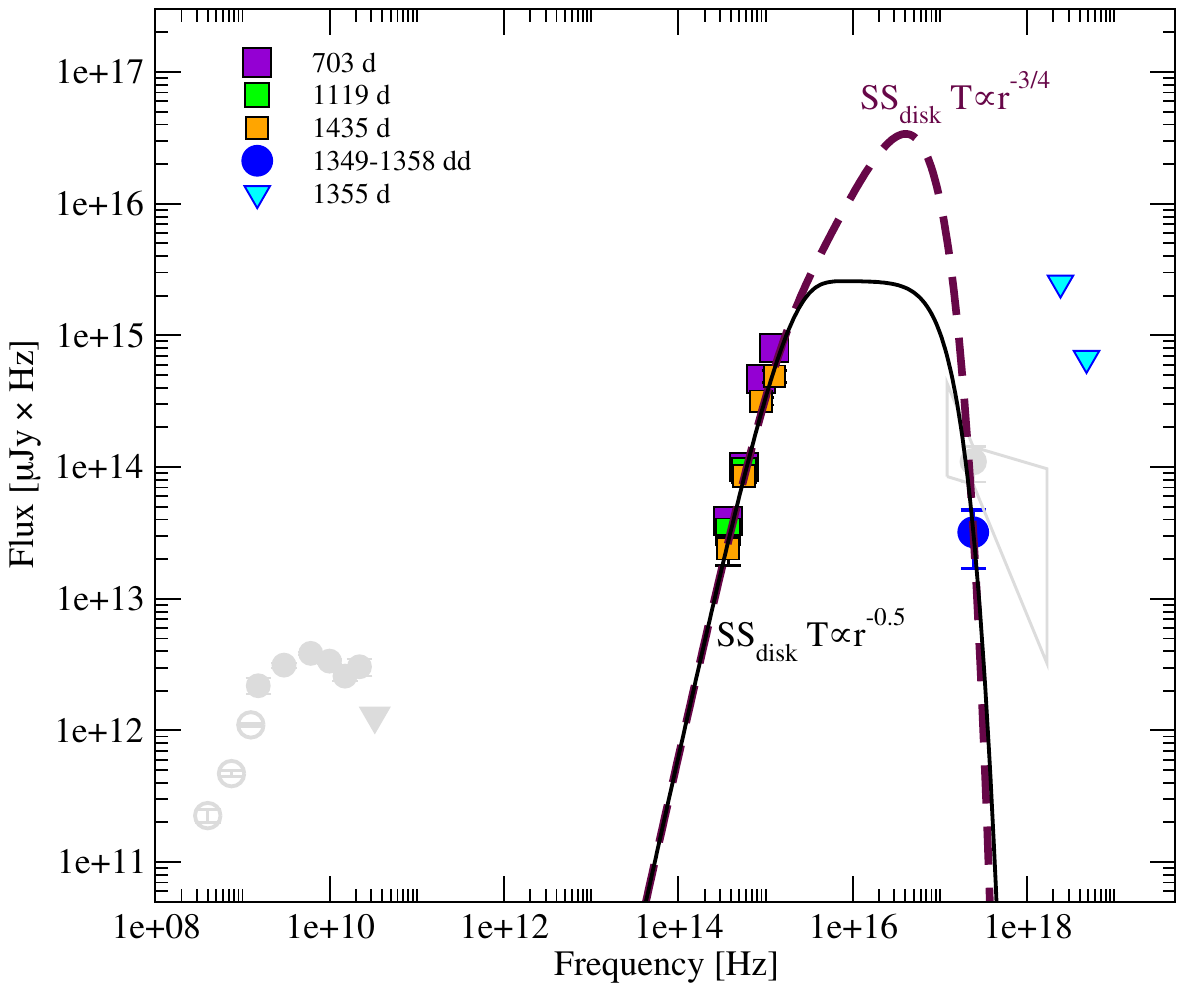}
\caption{\emph{Left panel}: \at\, radio-to-Xray SED at $\delta t\sim$217-234 d. The red solid line is the best fit broken power-law model for the radio data \citep[GMRT and VLA from][ and Coppejans et al. in prep., respectively]{Nayana21cow}. The bow-tie is the best-fit power-law model ($\Gamma\sim$2.9) from the X-ray spectral analysis of Obs3. Grey points are data taken at $\delta t\sim$1100-1440 d.
\emph{Right panel}: optical-UV-to-X-ray SED  at $\delta t\sim$700-1440 d. While for the modeling we formally associate the X-ray flux of Obs4 with \at{}, this assumption is relaxed in the discussion on the nature of the accretor (see the text).
The optical-UV fluxes (colored squares) are the HST data from \citet{Chen2023a,Chen2023b}. Triangles are \nustar\, data. Grey points are data acquired at $\delta t\sim$217-234 d.
In both panels, the long-dashed line is a standard Shakura-Sunyaev ($T\propto r^{-3/4}$) disk modeling the HST and soft X-rays (blue dot) at $\delta t\sim$1350-1435 d, while  the thin black line assumes  a disk temperature evolution as $T\propto r^{-1/2}$ (i.e., \citealt{Shakura&Sunyaev73} ``windy'' solution, or disks where radiation is trapped within the inflow and advected onto the accretor). In addition, at $\delta t\sim$ 220 d, we added a power-law ($\Gamma\sim 2.0$) with an exponential cut-off at $\approx$3 keV, mimicking a fading Comptonized component.
\label{fig:f3}}
\end{figure*}

%%%%%%%%%%%%%%%%%%%%%%%%%%%%%%%%%%%%%%%%%%%%%%%%%%%%%%%%%%%%%%%%%%%%%%%%%%%%%%%%%%%%%%%%%%%%%%%%%%%%%%%%%
\section{Late-time Broadband Radio-UV-X-ray SEDs} \label{sec:SED}
\at{}\, was observed with the the Karl G. Jansky Very Large Array (JVLA) at $\delta t=82-328$\,d  (PI Coppejans; \citealt{Margutti+19,Ho+19}, Coppejans et al., in prep.), with the Giant Metrewave Radio Telescope (GMRT)  extending to $\delta t=570\,$d (PI Nayana; \citealt{Nayana21cow}) and with HST in the time window $\delta t=50-1475$\,d (PIs Y. Chen, A. Filippenko, R. Foley, M. Drout, A. Levan; \citealt{Sun22,Sun22b,Chen2023b,Chen2023a,Inkenhaag23}).  In the following we model the multi-wavelength data sets that are approximately coeval with \xmm\, Obs3 and Obs4.

Multi-frequency radio observations were performed at $\delta$t$=$217--221\,d (VLA, 1.5 to 33 GHz; Coppejans et al. in prep.), and $\delta$t$=$227-234\,d (GMRT, at 0.4, 0.75 and 1.25 GHz; \citealt{Nayana21cow}), thus coeval with \xmm\, Obs3.  
The radio to X-ray SED is shown in Figure \ref{fig:f3}. We fit the radio $F_{\nu}$  data with a smoothed broken power-law model of the form:
\begin{equation}\label{eqn:sed_fit}
    F_{\nu}(\nu)=F_{\rm pk}\Bigg[\bigg(\frac{\nu}{\nu_{b}}\bigg)^{-\alpha_{r,1}/s}+\bigg(\frac{\nu}{\nu_{b}}\bigg)^{-\alpha_{r,2}/s}\Bigg]^{-s},
\end{equation}
which is typical of the synchrotron emission originating from the deceleration of astrophysical outflows associated with TDEs or SNe (e.g., \citealt{ChevalierFransson17}).  We find a best-fitting break frequency $\nu_{b}=$4.2$\,\pm\,$0.7 GHz, peak flux density of $F_{\rm pk}=2.4\pm 1.4$ mJy, optically thick slope $\alpha_{r,1} =-$2.0$\pm$0.3 and optically thin slope $\alpha_{r,2}=$0.7$\pm$0.3 (with a smoothing parameter $s=1.4\pm$0.9). The VLA data will be discussed in detail elsewhere. Here it is sufficient to note that the extrapolation of the radio to the X-ray band severely underpredicts the X-ray flux, confirming that the emission in the two energy bands is produced by two distinct components, as found for the previous epochs \citep{Margutti+19}.  Specifically, the shock-CSM interaction power that explains the radio data cannot account for the bright X-ray (and UV, see Figure \ref{fig:f3}) emission from \at{}. 

Late-time multi-band UV photometry was acquired with HST at $\delta t=714-1136$\,d and $\delta t=1475$\,d (Figure \ref{fig:f2}). The analysis of these observations  has been presented in \cite{Sun22,Sun22b,Chen2023b,Chen2023a,Inkenhaag23}. These studies agree on the presence of a luminous ($L_{\rm UV}\gtrsim 10^{39}\,\rm{erg\,s^{-1}}$), blue ($T_{\rm eff}\gtrsim 10^{4.6}\,\rm{K}$) and persistent (albeit slowly fading) UV source at the location of \at{}.\footnote{In the following analysis we  adopt the HST photometry values presented in \cite{Chen2023b,Chen2023a}. However, we note that consistent findings would be obtained by using the photometry presented in the other studies, without any significant impact on our major conclusions.} The UV spectrum is consistent with the $F_{\nu}\propto \nu^2$ expected for the Rayleigh-Jeans tail of an optically thick thermal spectrum. \cite{Chen2023b} find indications of a chromatic flux decay, with a flattening of the NUV spectrum in the last HST epoch, possibly pointing at cooling and expansion of the emitting region. The (slow) decay of the UV emission is not consistent with a stellar origin and suggests instead a physical association with \at{}. Interestingly, similarly luminous, hot and lingering UV emission has been detected in TDEs in the lower BH mass bin (i.e., $M_{\bullet}\lesssim 10^{6.5}\,\rm{M_{\sun}}$) years after the optical flare \citep{vanVelzen19,Mummery23}, and the similarity of the UV behavior of \at{} and TDEs (and its astrophysical implications) has been recently pointed out by \cite{Inkenhaag23}.

HST data at $\delta t=1475$\,d are approximately coeval with \xmm\, Obs4 and \emph{NuSTAR} data.\footnote{Based on the very slow evolution of the source at these epochs both in the UV and X-rays we do not expect that the $\Delta t\sim100$\,d difference between the acquisition time of these data sets will impact our major conclusions.   } Given the soft X-ray spectrum and the similar temporal behavior, we consider the possibility that the optical-UV and X-ray emission belong to the same radiative component, which we first model with the standard \cite{Shakura&Sunyaev73} multi-color black-body thin disk with a radial temperature scaling $T(r)\propto r^{-3/4}$ (solid line in Figure \ref{fig:f3}). The best-fitting parameters are an inner disk temperature $T_{\rm in}\sim 10^{5.91\pm0.02}\,\rm{K}$ (which is $\approx 0.06\,\rm{keV}$), outer-disk temperature $T_{\rm out}\sim 10^{4.45\pm 0.18}\,\rm{K}$, and inner disk radius $R_{\rm in}\sim 10^{10.3 \pm 0.1}\,\rm{cm}$, which corresponds to the ISCO of a non-rotating BH with mass $M_{\bullet}\sim 2.4\times 10^4\,\rm{M_{\odot}}$ ($R_{\rm{ISCO}}=6R_{\rm g} $ and the gravitational radius is $R_{\rm g} \equiv GM_{\bullet}/c^{2} \approx 1.5\times 10^{6} (M_{\bullet}/10M_{\odot})$ cm). These parameters imply an outer disk radius  $R_{\rm out}\approx 37\,\rm{R_{\odot}}$  from which most of the UV radiation is generated, and a BH accretion rate well below the Eddington limit $\dot M_{\bullet}\approx 0.01 \dot M_{\rm{Edd}}$, where we follow the convention $\dot M_{\rm{Edd}}\equiv L_{\rm{Edd}}/0.1\,c^2$. Less steep disk temperature profiles $T(r)\propto r^{-\alpha}$ with $\alpha<3/4$ such as those expected when radial flux advection is important (e.g., \citealt{Narayan08}) lead to similar inferences on $T_{\rm in}$, $T_{\rm out}$ and $\dot M/ \dot M_{\rm{Edd}}$, but smaller accretor masses $10^3\,\rm{M_{\odot}}\lesssim M_{\bullet}\lesssim 10^4\rm{M_{\odot}}$ (Figure \ref{fig:f3}).

The inferred low temperature $T_{\rm in}\sim T_X$ of the inner disk, set to truncate at the ISCO, is what drives the inference of IMBH-like accretor masses accreting at sub-Eddington rates (parenthetically, a similar argument motivated IMBH models for ULXs, e.g., \citealt{Miller04a,Miller04b,King23}). For $\dot{M}_{\bullet} \lesssim \dot{M}_{\rm Edd}$, the disk will be a standard thin disk (e.g., \citealt{Frank+02}), for which the effective temperature of the radiation from the innermost radii is approximately given by:
\begin{equation}
\begin{split}
kT_{\rm X} \approx k\left(\frac{3GM_{\bullet}\dot{M}}{8\pi \sigma R_{\rm ISCO}^{3}}\right)^{1/4} \simeq \\ 0.1\,{\rm keV}\,
\left(\frac{\dot{M}_{\bullet}}{0.01\,\dot{M}_{\rm Edd}}\right)^{1/4}\left(\frac{M_{\bullet}}{10^4\,M_{\odot}}\right)^{-1/4}
\left(\frac{R_{\rm ISCO}}{4R_{\rm g}}\right)^{-3/4}.
\label{eq:Tthin}
\end{split}
\end{equation}

Within this  set of models, from Eq. \ref{eq:Tthin},  smaller-accretors with $M_{\bullet}\sim 10-100 \,\rm{M_{\odot}}$ are coupled with  inner disks that are hotter than (and not consistent with) the observed X-rays. We note that this argument is independent from the detected X-rays being physically associated with \at{} or instead just providing a limit on the emission from \at{}. Mathematically, this is a consequence of $\nu F_{\nu}|_{UV}> \nu F_{\nu}|_{X}$. For the same reason, our inferences are consistent with those of \cite{Chen2023b} where the authors used a shallower X-ray limit from \emph{Swift}-XRT observations to reach a similar conclusion. However, differently from previous studies, we argue that these geometrically thin-disk models  cannot be used to rule out $M_{\bullet}\sim 10-100 \,\rm{M_{\odot}}$ BH accretors as in that case super-Eddington mass supplies $\dot{M}_{\bullet} \gg \dot{M}_{\rm Edd}$ are needed, which causes strong deviations from the geometrically thin-disk picture  since $H/R \sim L(R)/L_{\rm{Edd}}$, where $H$ and $R$ are the disk height and radius, respectively (e.g., \citealt{Frank+02}).  We discuss models of super-critical accretion that might apply to LFBOTs in \S\ref{sec:disc2}.

In this scenario, the disk is supposed to be present also at earlier times. Modelling of the optical-UV to X-ray emission at $\delta t=217-703$\,d provides us with parameter values consistent with those of the last epoch, with the caveat that the optical-UV flux at the time of Obs3 could have been higher. Interestingly, this model leaves an excess in the X-ray spectrum of Obs3, as supported by the X-ray spectral fit of a black body model resulting in a higher temperature and lower normalization (see \S\ref{sec:Xrayfit}. One possibility is that at $\delta t=218$\,d we are still detecting the non-thermal component, which was dominant at early times and is now rapidly fading (see Figure \ref{fig:f3}) and possibly driving the fast decay of the X-ray lightcurve at $\delta t>24$\,d.

%%%%%%%%%%%%%%%%%%%%%%%%%%%%%%%%%%%%%%%%%%%%%%%%%%%%%%%%%%%%%%%%%%%%%%%%%%%%%%
\section{Discussion}\label{sec:disc2}
Deep late-time \xmm\, observations out to $\delta t\approx 1360\,$d revealed a distinctive temporal and spectral evolution (Fig. \ref{fig:f2}) of the X-ray emission from the direction consistent with \at. Additionally, the last two \xmm\, observations of \at\, at $\delta t>200\,$d indicate a spectral change from a relatively hard to a very soft spectrum. Given that most of the X-ray photons are detected at very soft X-ray energies, obscuration is unlikely. We note that the lack of an intrinsic absorber is in line with observations at early times (\S\ref{sec:XMM}).

We discuss the implications of a soft X-ray source of emission in a LFBOT and its connection with the luminous, declining UV source, reported by \cite{Sun22b} and \cite{Chen2023b}, within the context of the shock interaction model (\S\ref{SubSec:interaction2}) and accretion-powered emission on a compact object (\S\ref{SubSec:CompactObject2}). In the following, we assume the association of the X-ray emission with \at{} robust for Obs3. Our main arguments hold true independently from the soft X-ray source of Obs4 being physically associated with \at{} or regarded as a limit. 

%%-----------------------------------------------------------------------------------------------
\subsection{Shock Interaction with the Medium} \label{SubSec:interaction2}
The SED of \at{} at $\sim$220 days clearly shows that the emission in the radio and X-ray bands stems from two separate radiative components and that the X-ray emission is in clear excess to the extrapolation of the synchrotron spectrum that best fits the shock-interaction-powered radio emission of \at~ (Figure \ref{fig:f2}, lower panel). However, for shocks that propagate within regions of CSM with high densities the X-ray emission is expected to be dominated by thermal bremsstrahlung radiation (e.g., \citealt{Fransson96,ChevalierFransson17}), as it was observationally confirmed in SN\,2014C \citep{Margutti14C,Thomas22,Brethauer22}. In this scenario, and assuming electron-ion equipartition, the temperature of the emission $T$ is related to the shock velocity as $T\approx 2.27\times 10^9 \mu_p v_4^2\, \rm{K}$, where $v_4$ is the shock velocity in units of $10^4\,\rm{km\,s^{-1}}$, $\mu_p\approx 0.6$ is the mean mass per particle including protons and electrons and we assumed solar composition and complete ionization, which is appropriate for a forward-shock powered X-ray emission.\footnote{For shocked SN ejecta chemical composition and complete ionization, which is appropriate for a reverse-shock powered X-ray emission, $\mu_p\approx 1.33$. } The observed very soft spectrum implies $k_bT\ll 0.2$ keV or $v\ll 250\,\rm{km\,s^{-1}}$, which is in stark contrast with the $v\sim0.2c$ derived from the radio modeling at 257\,d \citep{Nayana21cow}. We conclude that the forward shock that is powering the radio emission cannot be the source of energy behind the persistent, soft X-ray emission. A similar conclusion is supported by the independent analysis of the late-time UV emission \citep{Chen2023b} that we do not replicate here. 

%%-----------------------------------------------------------------------------------------------
\subsection{Accretion-powered scenarios for \at~ and LFBOTs} \label{SubSec:CompactObject2}
In this section we explore scenarios that connect the observed late-time UV-X-rays with manifestation of accretion processes onto a compact object. We start by summarizing the inferences from the early time data at $ \delta t<200$\,d that are relevant here and independently support this physical scenario.  

Generically, Compton-hump spectra like those observed in \at{} (\citealt{Margutti+19}, their Figure 6) have  been observationally associated with accreting sources like those of Active Galactic Nuclei (AGN) or X-Ray Binaries (XRBs), e.g, \cite{Reynolds99,Risaliti13,Belloni16}. More specifically: (i) the observed X-ray and UV emission from \at{} reaching  $L_{x+UV}\approx 5\times 10^{44}\,\rm{erg\,s^{-1}}$, if accretion powered, implies a super-Eddington regime for any accretor with mass $\le 10^6\,\rm{M_{\sun}}$. Borrowing the argument from the ULX literature (e.g., \citealt{King23}), this large luminosity, if disk powered, requires beaming of the emission even for super-Eddington mass supplies. (ii) This super-critical accretion scenario is independently supported by the significantly blue-shifted Fe K$\alpha$ fluorescent line  observed in \at{} at $\delta t\le 8$\,d \citep{Margutti+19}.  The numerical simulations by \citet{Thomsen2019}, which were originally motivated by the discovery of such a strongly blue-shifted Fe-K$\alpha$ line in the jetted relativistic TDE Swift~J1644 \citep{Kara16}, showed that hot (6.7--6.97 keV rest-frame), strongly blue-shifted ($\approx$8 keV centroid) Fe K$\alpha$ fluorescent lines are a robust observational signature of super-Eddington accretion disks irradiated by a lamp-post corona. Optically thick outflows are launched from the super-Eddington disk and create a conical funnel structure that provides the reflecting surface shaping the line profile, in close similarity to the model invoked by \cite{Margutti+19}. Note that at $ \delta t=218$\,d, we could still be detecting the rapidly cooling corona emission. (iii) The outflow velocity required to produce the observed Fe K$\alpha$ line shift is $v\approx$\,0.2$c$, which is similar to the inferred velocity of the optical photosphere of \at{} at early times, and to the velocity of its radio emitting  material \citep{Ho+19,Margutti+19,Nayana21cow}.

These velocities are significantly larger than thermal (or magneto-centrifugal) winds from sub-Eddington XRBs, but intriguingly similar to the velocity  of $\approx$ a few $0.1$c expected to be associated with outflows launched by super-Eddington accretion disks (e.g., \citealt{Thomsen2019} and references therein).\footnote{Consistently, these ``ultra-fast outflows'' have been located in several ULXs, see e.g. \cite{King23} for a recent review.}

%%----------------------------------------------------
\subsubsection{Inner accretion disk temperature: IMBHs vs. $M_{\bullet}\sim10-100\,\rm{M_{\odot}}$ BHs} 
\label{SubSubSec:InnerTe}
The inferences above support a physical connection between the early emission from \at{}\, and the manifestation of a super-Eddington accretion phase (like in jetted TDEs, GRBs and ULXs) independently from the accretor mass. From an X-ray perspective, the X-ray constraints at $ \delta t\gtrsim 200$\,d indicate that the system evolves from  luminous, spectrally hard and super-Eddington regime, to remarkably fainter, softer and $\le$ Eddington one (Figure \ref{fig:f2}). We thus consider that one of three potential transient events (i.e., a massive stellar core collapse, Wolf Rayet-BH merger, IMBH-TDE) results in the formation of a (super-Eddington) accretion disk of initial outer radius $R_{\rm d,0}$ around a black hole of mass $M_{\bullet}$.  The disk spreads with time as $R_{\rm d} (t)$ and evolves to an innermost ring temperature $T_X$ by the time of our  X-ray observations. 

For $\dot{M}_{\bullet} \lesssim \dot{M}_{\rm Edd} \equiv L_{\rm Edd}/(0.1c^{2})$, the disk will be a standard thin-disk, for which the effective temperature of the radiation from the innermost radii is approximately given by Equation \ref{eq:Tthin}. However, for $\dot{M}_{\bullet} \gg \dot{M}_{\rm Edd}$ we expect powerful outflows from the inner regions of the disk (e.g., \citealt{Narayan&Yi95,Blandford&Begelman99,Kitaki+21}), which can carry a substantial fraction of accreted matter and that can place the photosphere radius at radii $\gg R_{\rm ISCO}$.  In particular, a steady quasi-spherical wind with a mass-loss rate $\dot{M}_{\rm w} \sim \dot{M}_{\bullet}$ and velocity $v_{\rm w} \sim c$ launched from $\sim R_{\rm ISCO}$ will possess a density profile at radii $r \gg R_{\rm ISCO}$ given by $\rho_{\rm w} = \frac{\dot{M}_{\rm w}}{4\pi r^{2} v_{\rm w}}$.

The optical depth of the wind above a radius $r$ is
\begin{equation}
\tau_{\rm w} = \int_{r}^{\infty} \rho_{\rm w}\kappa dr = \frac{\dot{M}_{\rm w}\kappa}{4\pi r v_w},
\end{equation}
such that the photosphere radius ($\tau_{\rm w} =1$) is
\begin{equation}
r_{\rm ph} = \frac{\kappa\dot{M}_{\rm w}}{4\pi v_w} = 10R_{\rm g}\left(\frac{\dot{M}_{\rm w}}{\dot{M}_{\bullet}}\right)\left(\frac{\dot{M}_{\bullet}}{\dot{M}_{\rm Edd}}\right)\left(\frac{v_{\rm w}}{c}\right)^{-1},
\end{equation}
where we take an electron-scattering dominated opacity (for fully ionized gas with H-dominated composition) $\kappa\approx \sigma_T/m_p\approx 0.38\,$cm$^{2}$ g$^{-1}$, $\sigma_T$ and $m_p$ being the Thompson cross section and the proton mass, respectively.
  Assuming the emitted luminosity follows $L_{\rm X} \simeq \eta \dot{M}_{\bullet}c^{2}$ with radiative efficiency $\eta \sim 0.03$ as predicted by simulations of super-Eddington accretion (e.g., \citealt{Sadowski&Narayan16}), this predicts a maximum emission temperature for thermal emission from the innermost radii of the disk given by
\begin{equation}
\begin{split}
kT_{\rm X} \approx \left(\frac{L_{\rm X}}{4\pi \sigma r_{\rm ph}^{2}}\right)^{1/4} \\ \approx 0.85\,{\rm keV}\,\eta_{-1}^{1/4}\left(\frac{M_{\bullet}}{10M_{\odot}}\right)^{-1/4}\left(\frac{\dot{M}_{\bullet}}{\dot{M}_{\rm Edd}}\right)^{-1/4}\left(\frac{v_{\rm w}}{c}\right)^{-1/2},
\label{eq:TsuperEdd}
\end{split}
\end{equation}
where $\eta_{-1} \equiv \eta/(0.1)$, we have taken $\dot{M}_{\rm w} \sim \dot{M}_{\bullet}$ as expected for super-Eddington accretion disks that lose an order-unity fraction of their inflowing mass across each decade of radius (e.g., \citealt{Blandford&Begelman99}) and we used $L_{\rm X} = \eta \dot{M}_{\bullet}c^{2}$. 

Unlike a thin disk for which $T_{\rm X}$ increases for higher accretion rate (Equation \ref{eq:Tthin}), for super-Eddington accretion the thermal photosphere temperature decreases for higher accretion rate.\footnote{Though we note that non-thermal X-ray emission such as that observed at early times in \at{}\, can originate above the photosphere and will not follow Equation \ref{eq:TsuperEdd}.} For example, if $M_{\bullet} = 100\,M_{\odot}$, $\dot{M}_{\bullet} \sim 10 \dot{M}_{\rm Edd}$ and $\eta \sim 0.01$, Equation \ref{eq:TsuperEdd} gives $kT_{\rm X} \approx 0.15\,\rm{keV}$  similar to the thin-disk temperature (Equation \ref{eq:Tthin}) for an higher-mass IMBH $M_{\bullet} \sim 10^{4}M_{\odot}$ accreting at $\dot{M}_{\bullet} \ll \dot{M}_{\rm Edd}$, corresponding to $L_{X} \sim 10^{39}\,\rm{erg\,s^{-1}}$ (similar to our last observations of \at{}, Fig. \ref{fig:f2}). Stated another way, for any observed $L_{\rm X}$ and $T_{\rm X},$ there can be two allowed solutions: sub-Eddington for a high-mass (i.e. IMBH-like) BH (Eq. \ref{eq:Tthin}), and super-Eddington for a low-mass BH (Eq. \ref{eq:TsuperEdd}).  This argument applies independently from the physical association of the late-time soft X-ray source to \at{}.

%%----------------------------------------------------
\subsubsection{Outer accretion disk radius: IMBHs vs. $M_{\bullet}\sim10-100\,\rm{M_{\odot}}$ BHs}
\label{SubSubSec:OutR}
Late-time UV observations of \at{}\, constrain the outer disk radius at $\delta t\approx 1500$\,d to a value $R_{\rm d}\approx 40\,\rm{R_{\sun}}$ (\S\ref{sec:SED}). In this section we discuss the expectations from accretion disks formed by a core-collapse stellar explosion, an IMBH-TDE and a WR-BH merger. 

In the core-collapse case, the initial  disk radius  $R_{\rm d,0}$ depends on the angular momentum of the progenitor star, but$-$insofar that angular momentum is at a premium (e.g., \citealt{Fuller&Ma19})$-$is typically expected to be just outside the ISCO radius of at most tens of gravitational radii $R_{\rm g} \equiv GM_{\bullet}/c^{2} \approx 1.5\times 10^{6} (M_{\bullet}/10M_{\odot})$ cm.  So, we expect $R_{\rm d,0} \lesssim 10^{7}$ cm in the core-collapse case.  

In the IMBH-TDE case, the characteristic disk radius is at most (i.e., for a typical $\beta \simeq 1$ encounter, where $\beta\equiv R_{\rm t}/R_{\rm p}$ is the penetration parameter, $R_{\rm p}$ is the periapse) equal to twice the tidal radius $R_{\rm t}$:
\begin{equation}
\begin{split}
R_{\rm d,0} \simeq 2R_{\rm t} \simeq 2R_{\star}\left(\frac{M_{\bullet}}{M_{\star}}\right)^{1/3} \\
\approx 3\times 10^{12}\,{\rm cm}\,\left(\frac{R_{\star}}{R_{\odot}}\right)\left(\frac{M_{\star}}{M_{\odot}}\right)^{-1/3}\left(\frac{M_{\bullet}}{10^{4}M_{\odot}}\right)^{1/3},
\label{eq:Rd0}
\end{split}
\end{equation}
where $M_{\star}$ and $R_{\star}$ are the mass and radius of the disrupted star, respectively, and we have normalized $M_{\bullet}$ to $10^{4}M_{\odot}$ to match the fall-back time of the debris $t_{\rm fb} \simeq 5.8\,{\rm d}(M_{\bullet}/10^{4}M_{\odot})^{1/2}$ (e.g., \citealt{Stone+13}) with the \at{}\, peak engine duration $\sim$ optical rise-time $t_{\rm rise} \sim$ 3 d.  

The disruption process setting the disk size in the WR merger case is similar to the TDE scenario, except the BH is much less massive $M_{\bullet} \sim 10-100\, M_{\odot}$ and the mass of the WR star $M_{\star} \sim 10\,M_{\odot}$ is generally greater than the Sun, resulting in $R_{\rm d,0} \sim R_{\odot} \sim 10^{11}$ cm (e.g., \citealt{Metzger22}). 

After forming, the disk will begin to accrete on the viscous timescale
\begin{equation}
\begin{split}
t_{\rm visc,0} \sim  \left.\frac{r^{2}}{\nu}\right|_{R_{\rm d,0}} \\ %\sim \frac{1}{\alpha}\frac{1}{\theta_0^{2}}\left(\frac{R_{\rm d,0}^{3}}{GM_{\bullet}}\right)^{1/2} \\
\approx 5.0\,{\rm d}\, \alpha_{-2}^{-1}\left(\frac{R_{\rm d,0}}{R_{\odot}}\right)^{3/2}\left(\frac{M_{\bullet}}{10M_{\odot}}\right)^{-1/2}\left(\frac{H/r}{1/3}\right)^{-2},  \label{eq:tvisc2}
\end{split}
\end{equation}
where $\nu = \alpha c_{\rm s}H = \alpha r^{2}\Omega_{\rm K}(H/r)^{2}$ is the effective kinematic viscosity, $\Omega_{\rm K} = (GM_{\bullet}/r^{3})^{1/2}$, $\alpha = 10^{-2}\alpha_{-2}$,  $c_{\rm s}$ is the sound speed and we have normalized the equation to the case of a geometrically thick disk with $H/r\sim 1/3$ (see below).  In the WR-BH merger case, the LFBOT engine timescale $\lesssim 3$ d (for $\alpha \sim 0.01-0.1$) is naturally set by $t_{\rm visc,0}.$   However, in the core-collapse case $R_{\rm d,0}\ll R_{\sun}$ and as a result $t_{\rm visc,0}$ is too short to represent the engine activity timescale of \at{}, which would instead be set by the free-fall time of the infalling stellar envelope.  The viscous time in the IMBH-TDE case can also (just barely) obey $t_{\rm visc,0} \lesssim t_{\rm rise}$ for high $\alpha$ and/or small stellar radius $R_{\star} \lesssim R_{\odot}$.

On timescales $t \gtrsim t_{\rm visc,0}$, the disk will establish a steady flow onto the central compact object.   Taking $M_{\rm d,0} \sim M_{\star} \sim 1-10M_{\odot}$ as the mass of the disk, the peak accretion rate near the outer disk edge $\sim R_{\rm d,0}$ can be estimated as,
\begin{equation}
\dot{M}_0 \sim \frac{M_{\rm d,0}}{t_{\rm visc,0}} \sim 10^{28}{\rm g\,s^{-1}}\, \left(\frac{M_{\rm d,0}}{M_{\odot}}\right)\left(\frac{t_{\rm visc,0}}{3\,{\rm d}}\right)^{-1}.
\label{eq:Mdot0}
\end{equation}
This is $\gtrsim 6-10$ orders of magnitude larger than the Eddington rate $\dot{M}_{\rm Edd} \equiv L_{\rm Edd}/(0.1c^{2}) \sim (M_{\bullet}/10M_{\odot})10^{19}$ g s$^{-1}$, justifying our assumption of a geometrically thick disk with $H/r \sim 1/3$ (e.g., \citealt{Sadowski&Narayan15}).  

For the high mass inflow rates $\dot{M} \gg \dot{M}_{\rm trap} \equiv 10\,\dot{M}_{\rm Edd}(R_{\rm d,0}/R_{\rm g}) \sim 10^{4} \dot{M}_{\rm Edd}$ of interest, photons are trapped and advected inwards through the disk at radii $\lesssim R_{\rm d,0}$ (e.g., \citealt{Begelman79}).  Since the disk cannot cool efficiently via radiation (e.g., \citealt{Shakura&Sunyaev73}), the accretion flow in this ``hyper-accretion'' regime is susceptible to outflows powered by the released gravitational energy (e.g., \citealt{Narayan&Yi95,Blandford&Begelman99,Kitaki+21}). These outflows are responsible for creating the conical funnel geometry where the blue-shifted Fe K$\alpha$ line is formed  and provide a natural explanation to the \at{}\, fastest ejecta traced by radio observations (\S\ref{SubSec:CompactObject2}).  Such outflows cause the mass inflow rate $\dot{M}$ to decrease approaching the BH surface, in a way typically parametrized as a power-law in radius, 
\begin{equation}
\dot{M}(r) \approx \dot{M}_0 \left(\frac{r}{R_{\rm d,0}}\right)^{p},
\label{eq:Mdotr}
\end{equation}
where values for the parameter $p \approx 0.6$ are motivated by numerical simulations of radiatively inefficient accretion flows (e.g., \citealt{Yuan&Narayan14}). 

At late times $t \gg t_{\rm visc,0}$, the outer edge of the disk will spread outwards from its initial radius $R_{\rm d,0}$ due to the redistribution of angular momentum:
\begin{equation}
R_{\rm d}(t) \simeq R_{\rm d,0}\left(\frac{t}{t_{\rm visc,0}}\right)^{m}, t \gg t_{\rm visc,0},
\label{eq:Rd}
\end{equation}
where the parameter $m$ depends on the properties of the disk outflows.  If the disk outflows carry away only the local specific angular momentum of the disk material, then the outer edge of the disk will grow with time as $m = 2/3$ (e.g., \citealt{Cannizzo+90}).  However, in the case of a net torque on the disk, one can instead have $m = 1/(1.5+p) \simeq 0.48$ \citep{Metzger+08} where in the final equality we take $p = 0.6$.  

Thus, by the time of the HST observations $\delta t \sim 1500$\,d, the disk will spread to a radius $R_{\rm 1500\,d} \gtrsim (1500{\rm d}/t_{\rm visc,0})^{m}R_{\rm d,0} \sim 15-40\,R_{\rm d,0}$, where we demand $t_{\rm visc,0} \lesssim 3$ d to match the maximum initial rise time of \at{}.  Thus we predict $R_{\rm 1500\,d} \sim 15-40\,R_{\odot}$ for the WR-BH merger case, in overall agreement with the inferences from the late-time HST observations of \at{}, but generally significantly larger or smaller values in the IMBH-TDE or core-collapse cases, respectively. From this perspective, the WR-BH merger scenario provides a natural explanation of the inferred size of the accretion disk at late times, while the other two scenarios struggle explaining the size of the UV emitting region. We note that a potentially viable option is represented by the IMBH tidal disruption of a very small star with $R_{\star}\approx 0.1\,R_{\sun}$ and  $M_{\star}\approx 0.1\,M_{\sun}$. 

We end with a note on the expected evolution of the accretion power in the WR-BH merger and IMBH-TDE models. The accretion rate at disk radii $r < R_{\rm d}$ will drop as a power-law in time (e.g., \citealt{Metzger+08}), viz.~
\begin{equation}
\dot{M} \propto r^{p}t^{-4(p+1)/3}, t \gg t_{\rm visc,0},
\label{eq:Mdotlate}
\end{equation}
such that the accretion rate reaching the central BH will decay as
\begin{eqnarray}
\label{eq:MdotBH}
\dot{M}_{\bullet} \sim \dot{M}_0\left(\frac{R_{\rm g}}{R_{\rm d,0}}\right)^{p}\left(\frac{t}{t_{\rm visc,0}}\right)^{-\frac{4(p+1)}{3}}, %t \gg t_{\rm visc,0}, \nonumber \\
\end{eqnarray}
for $t \gg t_{\rm visc,0}$. This results in a jetted accretion power
\begin{eqnarray}
\begin{split}
L_{\rm acc} \sim \eta \dot{M}_{\bullet}c^{2}\approx \\
\approx 10^{45}\eta_{-2} \left(\frac{M_{\bullet}}{100\,\rm{M_{\sun}}}\right)^{0.4}
\left(\frac{M_{\rm d,0}}{10\,\rm{M_{\sun}}} \right)
\left(\frac{M_{\star}}{10\,\rm{M_{\sun}}} \right)^{0.2}\times\\
\times \left(\frac{R_{\star}}{1\,\rm{R_{\sun}}} \right)^{-0.6}
\left(\frac{t_{\rm visc,0}}{3\,\rm{day}} \right)^{1.1}
\left(\frac{t}{3\,\rm{day}} \right)^{-2.1}\,\,\rm{erg\,s^{-1}}
\\
%2\times 10^{44}\,{\rm erg\,s^{-1}}\eta_{-2}...\left(\frac{t}{\rm 3\,day}\right)^{-2.1},\,\,\, t \gtrsim t_{\rm visc,0}, 
\label{eq:Lacc}
\end{split}
\end{eqnarray}
where we have used Eqs.~(\ref{eq:Rd0}), (\ref{eq:Mdot0}), (\ref{eq:MdotBH}) for $p=0.6$. This expression is  valid for $t \gtrsim t_{\rm visc,0}$. We observe that $L_{\rm acc}$ is broadly similar in normalization and power-law decay rate $L_{\rm acc}\propto t^{-2.1}$ to the optical-UV light curve of \at{}\, (e.g., \citealt{Margutti+19}, their fig. 9; \citealt{Chen2023b}), suggesting that the early UV/optical emission from \at{}\, might also be powered by accretion processes onto a BH.

%%----------------------------------------------------
\subsection{Comparison with other LFBOTs: a unifying model } 
\label{SubSec:otherFBOTs}
The long-term evolution of the X-ray emission from the location of  \at\, is characterized by three regimes: (i) an initial, highly-luminous  slowly-decreasing phase at $\delta t\lesssim$24\,d with fast temporal variability superimposed, (ii) a rapid decay (about three order of magnitude drop in $\sim$200 days), possibly followed by (iii) a late-time flattening phase with $L_x\sim 4\times 10^{38}\,\rm{erg\,s^{-1}}$ at $\delta t\gtrsim$200\,d with a soft X-ray spectrum, which, however, might be contaminated, or even dominated, by the host emission. The presence of  luminous, lingering optical/UV emission during phase (iii) \citep{Sun22,Sun22b,Chen2023b,Chen2023a}, which is reminiscent of TDEs in the lowest mass bin \citep{vanVelzen19,Mummery23,Inkenhaag23}, indirectly supports the physical association of the persistent, soft X-ray emission with \at{}. 
 Indeed, a further deep X-ray observation, able to probe the flux level measured in Obs4, could help to better discriminate the \at{} contribution from the host galaxy one. 
 
The other five X-ray detected LFBOTs showed a  X-ray luminosity comparable to or larger than \at{}\, (Figure \ref{fig:f2}), with AT\,2022tsd rivaling the brightest long gamma-ray burst ever detected \citep{Matthews23}.  In the accretion-reprocessing scenario, more luminous X-ray emission can be a consequence of more powerful accretion and/or more favourable geometry \citep[i.e. a pole-on viewing angle,][]{Metzger22}, so that brighter or dimmer X-rays detected for other LFBOTs might be a consequence of the geometry of the emission and the location of the observer with respect to the pole.\footnote{We note that the highly aspherical geometry invoked here is consistent with the large polarization degree reaching $\sim7\%$ measured at optical wavelengths at a few days post-discovery of \at{} \citep{Maund23}.}
Since Fe K$\alpha$ fluorescent lines are visible only to non-equatorial observers (i.e. for observers at large angles from the poles the Fe K$\alpha$ emission is hidden by the optically thick disk and wind), it is plausible that the other LFBOTs with luminous X-rays also had transient Fe K$\alpha$ spectroscopic features (and Compton humps) that were missed because their X-ray monitoring started too late (Fig. \ref{fig:f2}). As a reference, the Fe K$\alpha$ line subsided at $\delta t>10$\,d in \at{}, and no other LFBOT has X-ray observations at these early epochs.  Similarly, the large distances of all the other LFBOTs prevent  meaningful constraints on the presence of faint persistent X-ray emission at late time $\delta t>200$\,d.  Therefore, other LFBOTs need to be discovered in the local universe (i.e. within $d\lesssim 50$\,Mpc) and at very early stages of their evolution (i.e. within days), in order to assess the occurrence of the Fe K$\alpha$ line complex, (and the associated Compton hump feature), and validate the late time soft X-ray flattening.
%Both the Fe K$\alpha$ line complex, (and the associated Compton hump feature), and the soft X-ray plateau will thus remain unique observational traits of \at{}\, until other LFBOTs will be discovered in the local universe (i.e. within $d\lesssim 50$\,Mpc) and at very early stages of their evolution (i.e. within days). 

We end by noting that this LFBOT unification model where the optical/UV and X-ray emission is powered by accretion (in the super-Eddington regime at early times, decreasing to $\le$ Eddington at late times), and the radio emission is powered by the deceleration of the outflows into the CSM, naturally predicts the presence of \at{}-like transients that are viewed edge-on, and will thus appear X-ray dim,  \emph{if} their optical/UV emission is detectable by these ``equatorial'' observers. %However, \raf{Comment on the optical/UV and radio: my guess is that the early UV will also appear dimmer for an edge-on observer for the same reason as the X-rays. Radio is less clear, but if the fast outflow with velocity of few 0.1c is within a funnel, it might be as well but less than the X-rays. Brian and others, what do you think? }

%%%%%%%%%%%%%%%%%%%%%%%%%%%%%%%%%%%%%%%%%%%%%%%%%%%%%%%%%%%%%%%%%%%%%%%%%%%%%%
\section{Summary and conclusions} \label{Sec:SummaryConclusions}
We presented the first deep X-ray observations of an LFBOT  at %the unprecedented epoch of 
$\sim 3.7$\,yrs after optical discovery. These observations sample a pristine portion of the observational phase space of FBOTs, and revealed the presence of a luminous, persistent, soft X-ray source of emission at the location of \at{}. The association with \at{} of such emission appears robust at $\delta t\sim 220$\, and uncertain at $\delta t\sim 1350$\,d, given the potentially comparable level of the X-ray flux from the host galaxy. We modeled the soft X-ray emission in the broader context of the late-time panchromatic observations \at{}, which includes luminous and slowly fading UV emission that is reminiscent of that detected from TDEs in the lowest mass bin years after the stellar disruption, as recently pointed out by \cite{Inkenhaag23}. We find that these late-time panchromatic observations of \at{}\, are consistent with either sub-Eddington accretion on an IMBH ($M_{\bullet}\approx 10^3-10^4\,\rm{M_{\sun}}$) or with the  manifestation of  $\sim$ Eddington-like accretion processes on a lower-mass accretor, e.g., a BH of  $M_{\bullet}\approx 10-100\,\rm{M_{\sun}}$. While similar arguments have been used in the ULX literature based on the detection of soft X-rays from these systems,\footnote{Specifically we note how IMBHs have been invoked in the ULX literature for systems that later were revealed to host pulsars, e.g. \cite{King23} and references therein.} we note that this conclusion is independent from the physical association of the soft X-ray source with \at{} (i.e. the argument applies even if we interpret the soft X-ray emission as a limit on the X-ray luminosity output of \at{}\, at these late epochs).

Assuming that the UV emission originates from the outermost annuli of a spreading disk, we showed that, with the exception of very small stars, the IMBH-TDE model ($M_{\bullet}\approx 10^3-10^4\,\rm{M_{\sun}}$) struggles to explain the inferred accretion disk radius, in addition to lacking a natural explanation for the dense CSM with a steep cutoff  at $r\sim $a few $10^{16}\,\rm{cm}$ implied by the analysis of the radio observations of \at{} and other LFBOTs.  The detection of an X-ray QPO signature in \at{}, if interpreted as an orbital frequency in the accretion disk, furthermore implies $M_{\bullet}<850\,\rm{M_{\sun}}$. Nevertheless, if the IMBH-TDE is the correct physical scenario powering LFBOTs, we predict that other LFBOTs will be associated with larger disks at late times (e.g., cooler-late time emission). On the other hand, the WR-BH merger scenario that invokes smaller masses BHs, offers a natural source of dense CSM within  $r\sim $a few $10^{16}\,\rm{cm}$ in the form of WR mass loss and relic disk from the common-envelope phase \citep{TunaMetzger23}, and while being able to explain the compact size of the disk at late times of our monitoring. A similar scenario,  involving a neutron star (NS) spiraling-in inside the envelope of a massive star has been proposed by \cite{Soker+19} (see also \citealt{Grichener2023}).
%%%%%%

Most importantly, independently from the BH mass, our work provides support to the hypothesis that LFBOTs are accretion-powered transients, thus also indirectly supporting the picture where the early, luminous and short-lived UV-optical emission that gives LFBOTs their name originates at least in part from (partial) reprocessing of the high-energy X-rays by outflows launched by the super-Eddington accretion disk.  In this accretion-powered scenario the LFBOTs observational properties at early times represent electromagnetic manifestations of super-Eddington accreting systems, later transitioning into an Eddington to sub-Eddington accretion ratio over the time scale of a few hundreds of days.  In this perspective LFBOTs qualify as new laboratories for super-Eddington accretion physics. However, \at{}\, is still the only event for which meaningful observations and detailed studies of its early (i.e. $\delta t<1$ week) \emph{and} late-time  (i.e. $\delta t>$ years) evolution  can be performed thanks to its proximity.
Going forward, wide field-of-view UV missions like ULTRASAT \citep{Sagiv14} and UVEX \citep{Kulkarni21}  will fill up this observational gap by providing early UV detections of LFBOTs in the local Universe. 
Finally, if related to an IMBH-TDE, LFBOTs are expected to produce gravitational waves, observable in the future in the local Universe ($\sim$10-25 Mpc) using the LISA detector \citep[e.g.][and references therein]{Eracleous2019}.\\

\acknowledgments
The first two authors have equally contributed to the realization of the paper.
The authors thank the anonymous referee for constructive suggestions that improved the clarity of the manuscript.
This paper is partially based on observations obtained with XMM-Newton, an ESA science mission with instruments and contributions directly funded by ESA Member States and NASA.  This material is based upon work partially supported by the National Aeronautics and Space Administration under Grant/Contract/Agreement No. 80NSSC22K0898 (PI Margutti).
This research has made use of data obtained from the 4XMM XMM-Newton serendipitous source catalogue compiled by the 10 institutes of the XMM-Newton Survey Science Centre selected by ESA. G.M. acknowledges financial support from INAF mini-grant "The high-energy view of jets and transient" (Bando Ricerca Fondamentale INAF 2022).  R.M. acknowledges partial support by the National Science Foundation under Grant No. AST-2221789 and AST-2224255, by the Heising-Simons Foundation under grant \# 2021-3248.

\vspace{5mm}
\facilities{HST(STIS), Swift(XRT and UVOT), NuSTAR, XMM,CXO}


\software{astropy \citep{Astropy}, scipy  \citep{scipy}, XMM-Newton SAS, HEAsoft, XSPEC \citep{AranudXSPEC}, CIAO \citep{Fruscione2006}. 
          }


\begin{deluxetable*}{lcccc}[b!]
\tablecaption{\xmm~observations 
\label{tab:log}}
\tablecolumns{5}
\tablenum{1}
\tablewidth{0pt}
\tablehead{
\multicolumn{5}{c}{\bf Observation log}\\
 &Obs3 &\multicolumn{3}{c}{Obs4}\\
  \cline{3-5}
\colhead{} &
\colhead{ 0822580601} &
\colhead{ 0843550401} &
\colhead{ 0843550501} &
\colhead{ 0843550601}\\ 
&  &(Obs4a) &(Obs4b) &(Obs4c)
% \cline{3-5}
% &(Obs3) &\multicolumn{3}{c}{Obs4}
 }
\startdata
Date (YYYY-mm-dd)     &2019-01-20      &2022-02-24 &2022-03-01 &2022-03-05\\
MJD           &58503.2        &59634.4    &59639.1    &59643.6\\
Time since discovery (d)   &218  &1349  &1354 &1358\\
Duration (ks)  &56.4 &39.3  &44.6  &44.6  \\
Exposure (ks) &42.5(pn)      &31.2(pn)        &7.6(pn)     &2.2(pn)\\
              &54.2(MOS1)       &37.4(MOS1)  &14.0(MOS1)  &15.3(MOS1)\\
              &54.4(MOS2)       &37.4(MOS2)  &17.0(MOS2)  &23.4(MOS2)\\
\hline
%\multicolumn{5}{c}{}\\
\multicolumn{5}{c}{{\bf \at\ detection}}\\
&Obs3 &\multicolumn{3}{c}{Obs4}\\
 \cline{3-5}
0.2-4.5 keV net count-rate (c/s) &(9.4$\pm$3.2)$\times$10$^{-4}$ (pn)   &\multicolumn{3}{c}{(5.7$\pm$2.4)$\times$10$^{-4}$ (pn)}  \\
               &(2.5$\pm$1.4)$\times$10$^{-4}$ (MOS1)    &\multicolumn{3}{c}{(7.1$\pm$1.9)$\times$10$^{-4}$ (MOS1)}\\
               &(4.5$\pm$1.5)$\times$10$^{-4}$ (MOS2)   &\multicolumn{3}{c}{(2.7$\pm$1.3)$\times$10$^{-4}$ (MOS2)}\\
               &(11.3$\pm$3.0)$\times$10$^{-4}$  (tot.) &\multicolumn{3}{c}{(6.3$\pm$2.0)$\times$10$^{-4}$ (tot.)}\\
detection likelihood &10.8   &\multicolumn{3}{c}{9.1}\\           
0.3-10 keV unabs. flux (\ergcm)  &(4.5$^{+1.5}_{-1.4}$)$\times$10$^{-15}$     &\multicolumn{3}{c}{(1.0$\pm$0.4)$\times$10$^{-15}$}\\
0.3-10 keV luminosity (\ergs)  &(1.9$\pm$0.6)$\times$10$^{39}$     &\multicolumn{3}{c}{(4.3$\pm$1.7)$\times$10$^{38}$}\\
\hline
\enddata
%\tablenotetext{a}{}
\tablecomments{Detection count rates and likelihood have been obtained with {\tt\string edetect\_chain} (Obs3) and {\tt\string edetect\_stack} (Obs4), see \S\ref{sec:XMM}. The count rates are calculated under the entire PSF of the detection, corrected for background, vignetting, detector efficiency and gaps.
Unabsorbed fluxes have been derived from the spectral analysis for Obs3 and from the pn count rate using WebPIMMS for Obs4. In both cases, we used the spectral parameters from the best-fitting power-law model of Obs 3 ($\Gamma=$2.9 and \NHgal$=5\times 10^{20}$ cm$^{-2}$). }
\end{deluxetable*}

%%%%%%%%%%%%%%%%%%%%%%%%%%%%%%%%%%%%%%%%%%%%%%%%%%%%%%%%%%%%%%
\appendix
\section{Early-time Chandra observations }
\label{sec:Chandra}
{\it Chandra} observed \at~with the Low Energy Transmissions Gratings (LETG), using the High Resolution Camera (HRC)-Spectroscopy detectors for 48 ks, beginning on June 24, 2018 ($\delta t=$8.2\,d).  A quick reduction was presented in a telegram shortly after the data were taken \citep{2018gratingAtel}, and our more careful analysis is consistent with that work.  We use the TGCAT data reductions \citep{2011AJ....141..129H}, and use all orders of the spectrum for $\pm1$ to $\pm8$.  The initial goal for using this mode was to determine whether there was an extreme UV excess from this source, given the strong UV emission observed by
Swift-UVOT \citep{UVOTATel,Kuin18}.  Such an excess was not present, and the spectrum below about 0.2 keV was heavily dominated by background.  We thus fit the data between the chip gap energy and 7 keV.  For the negative orders, the range of fitted energy is 0.3-7 keV, while for the positive orders it is from 0.24-7 keV.

We group the data into bins with at least 100 counts, and fit using $\chi^2$ with \citet{Gehrels1986} weighting.  Using an absorbed power law spectrum, with the \texttt{Xspec} \citep{AranudXSPEC} \texttt{phabs} model within \texttt{Sherpa} \citep{Sherpa}, we find $N_H$ = $5.5^{+3.8}_{-2.7} \times 10^{20}$ cm$^{-2}$, thus consistent with \NHgal, $\Gamma$=1.66$^{+0.2}_{-0.17}$, and a total unabsorbed flux from 0.25-7.0 keV of $(8.5\pm0.7) \times10^{-12}$ erg sec$^{-1}$ cm$^{-2}$, with $\chi^2/\nu$ of 136/162.  Additional parameters are thus not well justified. The 0.3-10 keV luminosity is (4.8$\pm$0.4)$\times$10$^{42}$ \ergs, and it is in line with the values measured by the other X-ray observatories at the same epoch.

\bibliography{ms1,AT2018cow1,at2018FU,refs-1}

\providecommand{\noopsort}[1]{}
\begin{thebibliography}{}
\expandafter\ifx\csname natexlab\endcsname\relax\def\natexlab#1{#1}\fi
\providecommand{\url}[1]{\href{#1}{#1}}

\bibitem[{{Akashi} \& {Soker}(2021)}]{AS2021}
{Akashi}, M., \& {Soker}, N. 2021, \mnras, 501, 4053

\bibitem[{{Antoni} \& {Quataert}(2022)}]{Antoni22}
{Antoni}, A., \& {Quataert}, E. 2022, \mnras, 511, 176

\bibitem[{{Arcavi} {et~al.}(2016){Arcavi}, {Wolf}, {Howell}, {Bildsten},
  {Leloudas}, {Hardin}, {Prajs}, {Perley}, {Svirski}, {Gal-Yam}, {Katz},
  {McCully}, {Cenko}, {Lidman}, {Sullivan}, {Valenti}, {Astier}, {Balland},
  {Carlberg}, {Conley}, {Fouchez}, {Guy}, {Pain}, {Palanque-Delabrouille},
  {Perrett}, {Pritchet}, {Regnault}, {Rich}, \& {Ruhlmann-Kleider}}]{Arcavi16}
{Arcavi}, I., {Wolf}, W.~M., {Howell}, D.~A., {et~al.} 2016, \apj, 819, 35

\bibitem[{{Arnaud}(1996)}]{AranudXSPEC}
{Arnaud}, K.~A. 1996, in Astronomical Society of the Pacific Conference Series,
  Vol. 101, Astronomical Data Analysis Software and Systems V, ed. G.~H.
  {Jacoby} \& J.~{Barnes}, 17

\bibitem[{{Astropy Collaboration} {et~al.}(2013){Astropy Collaboration},
  {Robitaille}, {Tollerud}, {Greenfield}, {Droettboom}, {Bray}, {Aldcroft},
  {Davis}, {Ginsburg}, {Price-Whelan}, {Kerzendorf}, {Conley}, {Crighton},
  {Barbary}, {Muna}, {Ferguson}, {Grollier}, {Parikh}, {Nair}, {Unther},
  {Deil}, {Woillez}, {Conseil}, {Kramer}, {Turner}, {Singer}, {Fox}, {Weaver},
  {Zabalza}, {Edwards}, {Azalee Bostroem}, {Burke}, {Casey}, {Crawford},
  {Dencheva}, {Ely}, {Jenness}, {Labrie}, {Lim}, {Pierfederici}, {Pontzen},
  {Ptak}, {Refsdal}, {Servillat}, \& {Streicher}}]{Astropy}
{Astropy Collaboration}, {Robitaille}, T.~P., {Tollerud}, E.~J., {et~al.} 2013,
  \aap, 558, A33

\bibitem[{{Begelman}(1979)}]{Begelman79}
{Begelman}, M.~C. 1979, \mnras, 187, 237

\bibitem[{{Belloni} \& {Motta}(2016)}]{Belloni16}
{Belloni}, T.~M., \& {Motta}, S.~E. 2016, in Astrophysics and Space Science
  Library, Vol. 440, Astrophysics of Black Holes: From Fundamental Aspects to
  Latest Developments, ed. C.~{Bambi}, 61

\bibitem[{{Bietenholz} {et~al.}(2020){Bietenholz}, {Margutti}, {Coppejans},
  {Alexander}, {Argo}, {Bartel}, {Eftekhari}, {Milisavljevic}, {Terreran}, \&
  {Berger}}]{Bietenholz+2020}
{Bietenholz}, M.~F., {Margutti}, R., {Coppejans}, D., {et~al.} 2020, \mnras,
  491, 4735

\bibitem[{{Blandford} \& {Begelman}(1999)}]{Blandford&Begelman99}
{Blandford}, R.~D., \& {Begelman}, M.~C. 1999, \mnras, 303, L1

\bibitem[{{Brethauer} {et~al.}(2022){Brethauer}, {Margutti}, {Milisavljevic},
  {Bietenholz}, {Chornock}, {Coppejans}, {De Colle}, {Hajela}, {Terreran},
  {Vargas}, {DeMarchi}, {Harris}, {Jacobson-Gal{\'a}n}, {Kamble}, {Patnaude},
  \& {Stroh}}]{Brethauer22}
{Brethauer}, D., {Margutti}, R., {Milisavljevic}, D., {et~al.} 2022, \apj, 939,
  105

\bibitem[{{Bright} {et~al.}(2022){Bright}, {Margutti}, {Matthews}, {Brethauer},
  {Coppejans}, {Wieringa}, {Metzger}, {DeMarchi}, {Laskar}, {Romero},
  {Alexander}, {Horesh}, {Migliori}, {Chornock}, {Berger}, {Bietenholz},
  {Devlin}, {Dicker}, {Jacobson-Gal{\'a}n}, {Mason}, {Milisavljevic}, {Motta},
  {Mroczkowski}, {Ramirez-Ruiz}, {Rhodes}, {Sarazin}, {Sfaradi}, \&
  {Sievers}}]{Bright+22}
{Bright}, J.~S., {Margutti}, R., {Matthews}, D., {et~al.} 2022, \apj, 926, 112

\bibitem[{{Cannizzo} {et~al.}(1990){Cannizzo}, {Lee}, \&
  {Goodman}}]{Cannizzo+90}
{Cannizzo}, J.~K., {Lee}, H.~M., \& {Goodman}, J. 1990, \apj, 351, 38

\bibitem[{{Chen} {et~al.}(2023{\natexlab{a}}){Chen}, {Drout}, {Piro},
  {Kilpatrick}, {Foley}, {Rojas-Bravo}, \& {Magee}}]{Chen2023b}
{Chen}, Y., {Drout}, M.~R., {Piro}, A.~L., {et~al.} 2023{\natexlab{a}}, arXiv
  e-prints, arXiv:2303.03501

\bibitem[{{Chen} {et~al.}(2023{\natexlab{b}}){Chen}, {Drout}, {Piro},
  {Kilpatrick}, {Foley}, {Rojas-Bravo}, {Taggart}, {Siebert}, \&
  {Magee}}]{Chen2023a}
---. 2023{\natexlab{b}}, arXiv e-prints, arXiv:2303.03500

\bibitem[{{Chevalier} \& {Fransson}(2017)}]{ChevalierFransson17}
{Chevalier}, R.~A., \& {Fransson}, C. 2017, in Handbook of Supernovae, ed.
  A.~W. {Alsabti} \& P.~{Murdin}, 875

\bibitem[{{Chrimes} {et~al.}(2023){Chrimes}, {Jonker}, {Levan}, {Coppejans},
  {Gaspari}, {Gompertz}, {Groot}, {Malesani}, {Mummery}, {Stanway}, \&
  {Wiersema}}]{Chrimes23}
{Chrimes}, A.~A., {Jonker}, P.~G., {Levan}, A.~J., {et~al.} 2023, arXiv
  e-prints, arXiv:2307.01771

\bibitem[{{Coppejans} {et~al.}(2020){Coppejans}, {Margutti}, {Terreran},
  {et~al.}}]{Coppejans+20}
{Coppejans}, D.~L., {Margutti}, R., {Terreran}, G., {et~al.} 2020, \apjl, 895,
  L23

\bibitem[{{Dessart} {et~al.}(2022){Dessart}, {Hillier}, \&
  {Kuncarayakti}}]{Dessart+21interaction}
{Dessart}, L., {Hillier}, D.~J., \& {Kuncarayakti}, H. 2022, \aap, 658, A130

\bibitem[{{Drout} {et~al.}(2014){Drout}, {Chornock}, {Soderberg}, {Sanders},
  {McKinnon}, {Rest}, {Foley}, {Milisavljevic}, {Margutti}, {Berger},
  {Calkins}, {Fong}, {Gezari}, {Huber}, {Kankare}, {Kirshner}, {Leibler},
  {Lunnan}, {Mattila}, {Marion}, {Narayan}, {Riess}, {Roth}, {Scolnic},
  {Smartt}, {Tonry}, {Burgett}, {Chambers}, {Hodapp}, {Jedicke}, {Kaiser},
  {Magnier}, {Metcalfe}, {Morgan}, {Price}, \& {Waters}}]{Drout+2014}
{Drout}, M.~R., {Chornock}, R., {Soderberg}, A.~M., {et~al.} 2014, \apj, 794,
  23

\bibitem[{{Eracleous} {et~al.}(2019){Eracleous}, {Gezari}, {Sesana},
  {Bogdanovic}, {MacLeod}, {Roth}, \& {Dai}}]{Eracleous2019}
{Eracleous}, M., {Gezari}, S., {Sesana}, A., {et~al.} 2019, \baas, 51, 10

\bibitem[{{Fox} \& {Smith}(2019)}]{Fox&Smith19}
{Fox}, O.~D., \& {Smith}, N. 2019, \mnras, 488, 3772

\bibitem[{{Frank} {et~al.}(2002){Frank}, {King}, \& {Raine}}]{Frank+02}
{Frank}, J., {King}, A., \& {Raine}, D.~J. 2002, {Accretion Power in
  Astrophysics: Third Edition}

\bibitem[{{Fransson} {et~al.}(1996){Fransson}, {Lundqvist}, \&
  {Chevalier}}]{Fransson96}
{Fransson}, C., {Lundqvist}, P., \& {Chevalier}, R.~A. 1996, \apj, 461, 993

\bibitem[{{Freeman} {et~al.}(2001){Freeman}, {Doe}, \&
  {Siemiginowska}}]{Sherpa}
{Freeman}, P., {Doe}, S., \& {Siemiginowska}, A. 2001, in Society of
  Photo-Optical Instrumentation Engineers (SPIE) Conference Series, Vol. 4477,
  Astronomical Data Analysis, ed. J.-L. {Starck} \& F.~D. {Murtagh}, 76--87

\bibitem[{{Fruscione} {et~al.}(2006){Fruscione}, {McDowell}, {Allen},
  {Brickhouse}, {Burke}, {Davis}, {Durham}, {Elvis}, {Galle}, {Harris},
  {Huenemoerder}, {Houck}, {Ishibashi}, {Karovska}, {Nicastro}, {Noble},
  {Nowak}, {Primini}, {Siemiginowska}, {Smith}, \& {Wise}}]{Fruscione2006}
{Fruscione}, A., {McDowell}, J.~C., {Allen}, G.~E., {et~al.} 2006, in Society
  of Photo-Optical Instrumentation Engineers (SPIE) Conference Series, Vol.
  6270, Society of Photo-Optical Instrumentation Engineers (SPIE) Conference
  Series, ed. D.~R. {Silva} \& R.~E. {Doxsey}, 62701V

\bibitem[{{Fuller} \& {Ma}(2019)}]{Fuller&Ma19}
{Fuller}, J., \& {Ma}, L. 2019, \apjl, 881, L1

\bibitem[{{Gehrels}(1986)}]{Gehrels1986}
{Gehrels}, N. 1986, \apj, 303, 336

\bibitem[{{Gottlieb} {et~al.}(2022){Gottlieb}, {Tchekhovskoy}, \&
  {Margutti}}]{Gottlieb+2022}
{Gottlieb}, O., {Tchekhovskoy}, A., \& {Margutti}, R. 2022, \mnras, 513, 3810

\bibitem[{{Grichener}(2023)}]{Grichener2023}
{Grichener}, A. 2023, \mnras, 523, 221

\bibitem[{{Harrison} {et~al.}(2013){Harrison}, {Craig}, {Christensen},
  {Hailey}, {Zhang}, {Boggs}, {Stern}, {Cook}, {Forster}, {Giommi},
  {Grefenstette}, {Kim}, {Kitaguchi}, {Koglin}, {Madsen}, {Mao}, {Miyasaka},
  {Mori}, {Perri}, {Pivovaroff}, {Puccetti}, {Rana}, {Westergaard}, {Willis},
  {Zoglauer}, {An}, {Bachetti}, {Barri{\`e}re}, {Bellm}, {Bhalerao},
  {Brejnholt}, {Fuerst}, {Liebe}, {Markwardt}, {Nynka}, {Vogel}, {Walton},
  {Wik}, {Alexander}, {Cominsky}, {Hornschemeier}, {Hornstrup}, {Kaspi},
  {Madejski}, {Matt}, {Molendi}, {Smith}, {Tomsick}, {Ajello}, {Ballantyne},
  {Balokovi{\'c}}, {Barret}, {Bauer}, {Blandford}, {Brandt}, {Brenneman},
  {Chiang}, {Chakrabarty}, {Chenevez}, {Comastri}, {Dufour}, {Elvis}, {Fabian},
  {Farrah}, {Fryer}, {Gotthelf}, {Grindlay}, {Helfand}, {Krivonos}, {Meier},
  {Miller}, {Natalucci}, {Ogle}, {Ofek}, {Ptak}, {Reynolds}, {Rigby},
  {Tagliaferri}, {Thorsett}, {Treister}, \& {Urry}}]{Harrison13}
{Harrison}, F.~A., {Craig}, W.~W., {Christensen}, F.~E., {et~al.} 2013, \apj,
  770, 103

\bibitem[{{Ho} {et~al.}(2022{\natexlab{a}}){Ho}, {Perley}, {Chen}, {Schulze},
  {Sollerman}, \& {Gal-Yam}}]{Ho22opticalVar}
{Ho}, A.~Y.~Q., {Perley}, D.~A., {Chen}, P., {et~al.} 2022{\natexlab{a}},
  Transient Name Server AstroNote, 267, 1

\bibitem[{{Ho} {et~al.}(2020{\natexlab{a}}){Ho}, {Perley}, {Kulkarni}, {Dong},
  {et~al.}}]{Ho+20}
{Ho}, A. Y.~Q., {Perley}, D.~A., {Kulkarni}, S.~R., {Dong}, D. Z.~J., {et~al.}
  2020{\natexlab{a}}, \apj, 895, 49

\bibitem[{{Ho} {et~al.}(2019){Ho}, {Phinney}, {Ravi}, {Kulkarni}, {Petitpas},
  {Emonts}, {Bhalerao}, {Blundell}, {Cenko}, {Dobie}, {Howie}, {Kamraj},
  {Kasliwal}, {Murphy}, {Perley}, {Sridharan}, \& {Yoon}}]{Ho+19}
{Ho}, A. Y.~Q., {Phinney}, E.~S., {Ravi}, V., {et~al.} 2019, \apj, 871, 73

\bibitem[{{Ho} {et~al.}(2020{\natexlab{b}}){Ho}, {Perley}, {Kulkarni}, {Dong},
  {De}, {Chandra}, {Andreoni}, {Bellm}, {Burdge}, {Coughlin}, {Dekany},
  {Feeney}, {Frederiks}, {Fremling}, {Golkhou}, {Graham}, {Hale}, {Helou},
  {Horesh}, {Kasliwal}, {Laher}, {Masci}, {Miller}, {Porter}, {Ridnaia},
  {Rusholme}, {Shupe}, {Soumagnac}, \& {Svinkin}}]{Ho20AT2018lug}
{Ho}, A. Y.~Q., {Perley}, D.~A., {Kulkarni}, S.~R., {et~al.}
  2020{\natexlab{b}}, \apj, 895, 49

\bibitem[{{Ho} {et~al.}(2022{\natexlab{b}}){Ho}, {Perley}, {Yao}, {Svinkin},
  {de Ugarte Postigo}, {Perley}, {Kann}, {Burns}, {Andreoni}, {Bellm},
  {Bissaldi}, {Bloom}, {Brink}, {Dekany}, {Drake}, {Ag{\"u}{\'\i}
  Fern{\'a}ndez}, {Filippenko}, {Frederiks}, {Graham}, {Hristov}, {Kasliwal},
  {Kulkarni}, {Kumar}, {Laher}, {Lysenko}, {Mailyan}, {Malacaria}, {Miller},
  {Poolakkil}, {Riddle}, {Ridnaia}, {Rusholme}, {Savchenko}, {Sollerman},
  {Th{\"o}ne}, {Tsvetkova}, {Ulanov}, \& {von Kienlin}}]{Ho22sample}
{Ho}, A. Y.~Q., {Perley}, D.~A., {Yao}, Y., {et~al.} 2022{\natexlab{b}}, \apj,
  938, 85

\bibitem[{{Ho} {et~al.}(2022{\natexlab{c}}){Ho}, {Margalit}, {Bremer},
  {Perley}, {Yao}, {Dobie}, {Kaplan}, {O'Brien}, {Petitpas}, \& {Zic}}]{Ho+22}
{Ho}, A. Y.~Q., {Margalit}, B., {Bremer}, M., {et~al.} 2022{\natexlab{c}},
  \apj, 932, 116

\bibitem[{{Huenemoerder} {et~al.}(2011){Huenemoerder}, {Mitschang}, {Dewey},
  {Nowak}, {Schulz}, {Nichols}, {Davis}, {Houck}, {Marshall}, {Noble},
  {Morgan}, \& {Canizares}}]{2011AJ....141..129H}
{Huenemoerder}, D.~P., {Mitschang}, A., {Dewey}, D., {et~al.} 2011, \aj, 141,
  129

\bibitem[{{Inkenhaag} {et~al.}(2023){Inkenhaag}, {Jonker}, {Levan}, {Chrimes},
  {Mummery}, {Perley}, \& {Tanvir}}]{Inkenhaag23}
{Inkenhaag}, A., {Jonker}, P.~G., {Levan}, A.~J., {et~al.} 2023, arXiv
  e-prints, arXiv:2308.07381

\bibitem[{{Kalberla} {et~al.}(2005){Kalberla}, {Burton}, {Hartmann}, {Arnal},
  {Bajaja}, {Morras}, \& {P{\"o}ppel}}]{Kalberla05}
{Kalberla}, P.~M.~W., {Burton}, W.~B., {Hartmann}, D., {et~al.} 2005, \aap,
  440, 775

\bibitem[{{Kara} {et~al.}(2016){Kara}, {Miller}, {Reynolds}, \& {Dai}}]{Kara16}
{Kara}, E., {Miller}, J.~M., {Reynolds}, C., \& {Dai}, L. 2016, \nat, 535, 388

\bibitem[{{King} {et~al.}(2023){King}, {Lasota}, \& {Middleton}}]{King23}
{King}, A., {Lasota}, J.-P., \& {Middleton}, M. 2023, \nar, 96, 101672

\bibitem[{{Kitaki} {et~al.}(2021){Kitaki}, {Mineshige}, {Ohsuga}, \&
  {Kawashima}}]{Kitaki+21}
{Kitaki}, T., {Mineshige}, S., {Ohsuga}, K., \& {Kawashima}, T. 2021, arXiv
  e-prints, arXiv:2101.11028

\bibitem[{{Kuin} {et~al.}(2018){Kuin}, {Wu}, {Oates}, {Lien}, {Emery},
  {Kennea}, {de Pasquale}, {Han}, {Brown}, {Tohuvavohu}, {Breeveld}, {Burrows},
  {Cenko}, {Campana}, {Levan}, {Markwardt}, {Osborne}, {Page}, {Page},
  {Sbarufatti}, {Siegel}, \& {Troja}}]{Kuin18}
{Kuin}, N.~P.~M., {Wu}, K., {Oates}, S., {et~al.} 2018, ArXiv e-prints,
  arXiv:1808.08492

\bibitem[{{Kuin} {et~al.}(2019){Kuin}, {Wu}, {Oates}, {Lien}, {Emery},
  {Kennea}, {de Pasquale}, {Han}, {Brown}, {Tohuvavohu}, {et~al.}}]{Kuin+19}
{Kuin}, N. P.~M., {Wu}, K., {Oates}, S., {et~al.} 2019, \mnras, 487, 2505

\bibitem[{{Kulkarni} {et~al.}(2021){Kulkarni}, {Harrison}, {Grefenstette},
  {Earnshaw}, {Andreoni}, {Berg}, {Bloom}, {Cenko}, {Chornock}, {Christiansen},
  {Coughlin}, {Wuollet Criswell}, {Darvish}, {Das}, {De}, {Dessart}, {Dixon},
  {Dorsman}, {El-Badry}, {Evans}, {Ford}, {Fremling}, {Gansicke}, {Gezari},
  {Goetberg}, {Green}, {Graham}, {Heida}, {Ho}, {Jaodand}, {Johns-Krull},
  {Kasliwal}, {Lazzarini}, {Lu}, {Margutti}, {Martin}, {Masters}, {McKernan},
  {Naze}, {Nissanke}, {Parazin}, {Perley}, {Phinney}, {Piro}, {Raaijmakers},
  {Rauw}, {Rodriguez}, {Sana}, {Senchyna}, {Singer}, {Spake}, {Stassun},
  {Stern}, {Teplitz}, {Weisz}, \& {Yao}}]{Kulkarni21}
{Kulkarni}, S.~R., {Harrison}, F.~A., {Grefenstette}, B.~W., {et~al.} 2021,
  arXiv e-prints, arXiv:2111.15608

\bibitem[{{Lehmer} {et~al.}(2016){Lehmer}, {Basu-Zych}, {Mineo}, {Brandt},
  {Eufrasio}, {Fragos}, {Hornschemeier}, {Luo}, {Xue}, {Bauer}, {Gilfanov},
  {Ranalli}, {Schneider}, {Shemmer}, {Tozzi}, {Trump}, {Vignali}, {Wang},
  {Yukita}, \& {Zezas}}]{Lehmer2016}
{Lehmer}, B.~D., {Basu-Zych}, A.~R., {Mineo}, S., {et~al.} 2016, \apj, 825, 7

\bibitem[{{Leung} {et~al.}(2021){Leung}, {Fuller}, \& {Nomoto}}]{Leung+21}
{Leung}, S.-C., {Fuller}, J., \& {Nomoto}, K. 2021, \apj, 915, 80

\bibitem[{{Lyman} {et~al.}(2020){Lyman}, {Galbany}, {S{\'a}nchez}, {Anderson},
  {Kuncarayakti}, \& {Prieto}}]{Lyman+20}
{Lyman}, J.~D., {Galbany}, L., {S{\'a}nchez}, S.~F., {et~al.} 2020, \mnras,
  495, 992

\bibitem[{{Maccarone} {et~al.}(2018){Maccarone}, {Rivera Sandoval}, {Corsi},
  {Pooley}, \& {Knigge}}]{2018gratingAtel}
{Maccarone}, T.~J., {Rivera Sandoval}, L., {Corsi}, A.~r., {Pooley}, D., \&
  {Knigge}, C. 2018, The Astronomer's Telegram, 11779, 1

\bibitem[{{Margalit}(2022)}]{Margalit22}
{Margalit}, B. 2022, \apj, 933, 238

\bibitem[{{Margutti} {et~al.}(2019){Margutti}, {Metzger}, {Chornock},
  {et~al.}}]{Margutti+19}
{Margutti}, R., {Metzger}, B.~D., {Chornock}, R., {et~al.} 2019, \apj, 872, 18

\bibitem[{{Margutti} {et~al.}(2017){Margutti}, {Kamble}, {Milisavljevic},
  {Zapartas}, {de Mink}, {Drout}, {Chornock}, {Risaliti}, {Zauderer},
  {Bietenholz}, {Cantiello}, {Chakraborti}, {Chomiuk}, {Fong}, {Grefenstette},
  {Guidorzi}, {Kirshner}, {Parrent}, {Patnaude}, {Soderberg}, {Gehrels}, \&
  {Harrison}}]{Margutti14C}
{Margutti}, R., {Kamble}, A., {Milisavljevic}, D., {et~al.} 2017, \apj, 835,
  140

\bibitem[{{Matthews} {et~al.}(2023){Matthews}, {Margutti}, {Metzger},
  {Milisavljevic}, {Migliori}, {Laskar}, {Brethauer}, {Berger}, {Chornock},
  {Drout}, \& {Ramirez-Ruiz}}]{Matthews23}
{Matthews}, D.~J., {Margutti}, R., {Metzger}, B.~D., {et~al.} 2023, arXiv
  e-prints, arXiv:2306.01114

\bibitem[{{Maund} {et~al.}(2023){Maund}, {H{\"o}flich}, {Steele},
  {Yang(杨轶)}, {Wiersema}, {Kobayashi}, {Jordana-Mitjans}, {Mundell},
  {Gomboc}, {Guidorzi}, \& {Smith}}]{Maund23}
{Maund}, J.~R., {H{\"o}flich}, P.~A., {Steele}, I.~A., {et~al.} 2023, \mnras,
  521, 3323

\bibitem[{{Metzger}(2022)}]{Metzger22}
{Metzger}, B.~D. 2022, \apj, 932, 84

\bibitem[{{Metzger} \& {Perley}(2023)}]{Metzger23}
{Metzger}, B.~D., \& {Perley}, D.~A. 2023, \apj, 944, 74

\bibitem[{{Metzger} {et~al.}(2008){Metzger}, {Piro}, \&
  {Quataert}}]{Metzger+08}
{Metzger}, B.~D., {Piro}, A.~L., \& {Quataert}, E. 2008, \mnras, 390, 781

\bibitem[{{Micha{\l}owski} {et~al.}(2019){Micha{\l}owski}, {Kamphuis},
  {Hjorth}, {Kann}, {de Ugarte Postigo}, {Galbany}, {Fynbo}, {Ghosh}, {Hunt},
  {Kuncarayakti}, {Le Floc'h}, {Le{\'s}niewska}, {Misra}, {Nicuesa Guelbenzu},
  {Palazzi}, {Rasmussen}, {Resmi}, {Rossi}, {Savaglio}, {Schady}, {Schulze},
  {Th{\"o}ne}, {Watson}, {J{\'o}zsa}, {Serra}, \& {Smirnov}}]{Michalowski19}
{Micha{\l}owski}, M.~J., {Kamphuis}, P., {Hjorth}, J., {et~al.} 2019, \aap,
  627, A106

\bibitem[{{Miller} {et~al.}(2004{\natexlab{a}}){Miller}, {Fabian}, \&
  {Miller}}]{Miller04a}
{Miller}, J.~M., {Fabian}, A.~C., \& {Miller}, M.~C. 2004{\natexlab{a}}, \apjl,
  614, L117

\bibitem[{{Miller} {et~al.}(2004{\natexlab{b}}){Miller}, {Fabian}, \&
  {Miller}}]{Miller04b}
---. 2004{\natexlab{b}}, \apj, 607, 931

\bibitem[{{Mineo} {et~al.}(2014){Mineo}, {Gilfanov}, {Lehmer}, {Morrison}, \&
  {Sunyaev}}]{Mineo2014}
{Mineo}, S., {Gilfanov}, M., {Lehmer}, B.~D., {Morrison}, G.~E., \& {Sunyaev},
  R. 2014, \mnras, 437, 1698

\bibitem[{{Mineo} {et~al.}(2012){Mineo}, {Gilfanov}, \& {Sunyaev}}]{Mineo12}
{Mineo}, S., {Gilfanov}, M., \& {Sunyaev}, R. 2012, \mnras, 426, 1870

\bibitem[{{Mohan} {et~al.}(2020){Mohan}, {An}, \& {Yang}}]{Mohan+20}
{Mohan}, P., {An}, T., \& {Yang}, J. 2020, \apjl, 888, L24

\bibitem[{{Mummery} {et~al.}(2023){Mummery}, {van Velzen}, {Nathan}, {Ingram},
  {Hammerstein}, {Fraser-Taliente}, \& {Balbus}}]{Mummery23}
{Mummery}, A., {van Velzen}, S., {Nathan}, E., {et~al.} 2023, arXiv e-prints,
  arXiv:2308.08255

\bibitem[{{Narayan} \& {McClintock}(2008)}]{Narayan08}
{Narayan}, R., \& {McClintock}, J.~E. 2008, \nar, 51, 733

\bibitem[{{Narayan} \& {Yi}(1995)}]{Narayan&Yi95}
{Narayan}, R., \& {Yi}, I. 1995, \apj, 444, 231

\bibitem[{{Nayana} \& {Chandra}(2021)}]{Nayana21cow}
{Nayana}, A.~J., \& {Chandra}, P. 2021, \apjl, 912, L9

\bibitem[{{Nicholl} {et~al.}(2023){Nicholl}, {Srivastav}, {Fulton}, {Gomez},
  {Huber}, {Oates}, {Ramsden}, {Rhodes}, {Smartt}, {Smith}, {Aamer},
  {Anderson}, {Bauer}, {Berger}, {de Boer}, {Chambers}, {Charalampopoulos},
  {Chen}, {Fender}, {Fraser}, {Gao}, {Green}, {Galbany}, {Gompertz},
  {Gromadzki}, {Guti{\'e}rrez}, {Howell}, {Inserra}, {Jonker}, {Kopsacheili},
  {Lowe}, {Magnier}, {McCully}, {McGee}, {Moore}, {M{\"u}ller-Bravo},
  {Newsome}, {Gonzalez}, {Pellegrino}, {Pessi}, {Pursiainen}, {Rest}, {Ridley},
  {Shappee}, {Sheng}, {Smith}, {Terreran}, {Tucker}, {Vink{\'o}}, {Wainscoat},
  {Wiseman}, \& {Young}}]{Nicholl23}
{Nicholl}, M., {Srivastav}, S., {Fulton}, M.~D., {et~al.} 2023, \apjl, 954, L28

\bibitem[{{Pasham} {et~al.}(2021){Pasham}, {Ho}, {Alston}, {Remillard}, {Ng},
  {Gendreau}, {Metzger}, {Altamirano}, {Chakrabarty}, {Fabian}, {Miller},
  {Bult}, {Arzoumanian}, {Steiner}, {Strohmayer}, {Tombesi}, {Homan},
  {Cackett}, \& {Harding}}]{DJ21}
{Pasham}, D.~R., {Ho}, W. C.~G., {Alston}, W., {et~al.} 2021, Nature Astronomy,
  6, 249

\bibitem[{{Pellegrino} {et~al.}(2022){Pellegrino}, {Howell}, {Vink{\'o}},
  {Gangopadhyay}, {Xiang}, {Arcavi}, {Brown}, {Burke}, {Hiramatsu},
  {Hosseinzadeh}, {Li}, {McCully}, {Misra}, {Newsome}, {Gonzalez}, {Pritchard},
  {Valenti}, {Wang}, \& {Zhang}}]{Pellegrino+2022}
{Pellegrino}, C., {Howell}, D.~A., {Vink{\'o}}, J., {et~al.} 2022, \apj, 926,
  125

\bibitem[{{Perley} {et~al.}(2019){Perley}, {Mazzali}, {Yan},
  {et~al.}}]{Perley+19}
{Perley}, D.~A., {Mazzali}, P.~A., {Yan}, L., {et~al.} 2019, \mnras, 484, 1031

\bibitem[{{Poznanski} {et~al.}(2010){Poznanski}, {Chornock}, {Nugent}, {Bloom},
  {Filippenko}, {Ganeshalingam}, {Leonard}, {Li}, \& {Thomas}}]{Poznanski10}
{Poznanski}, D., {Chornock}, R., {Nugent}, P.~E., {et~al.} 2010, Science, 327,
  58

\bibitem[{{Prentice} {et~al.}(2018){Prentice}, {Maguire}, {Smartt}, {Magee},
  {Schady}, {Sim}, {Chen}, {Clark}, {Colin}, {Fulton}, {McBrien}, {O'Neill},
  {Smith}, {Ashall}, {Chambers}, {Denneau}, {Flewelling}, {Heinze}, {Holoien},
  {Huber}, {Kochanek}, {Mazzali}, {Prieto}, {Rest}, {Shappee}, {Stalder},
  {Stanek}, {Stritzinger}, {Thompson}, \& {Tonry}}]{Prentice+2018}
{Prentice}, S.~J., {Maguire}, K., {Smartt}, S.~J., {et~al.} 2018, \apjl, 865,
  L3

\bibitem[{{Pursiainen} {et~al.}(2018){Pursiainen}, {Childress}, {Smith},
  {Prajs}, {Sullivan}, {Davis}, {Foley}, {Asorey}, {Calcino}, {Carollo},
  {Curtin}, {D'Andrea}, {Glazebrook}, {Gutierrez}, {Hinton}, {Hoormann},
  {Inserra}, {Kessler}, {King}, {Kuehn}, {Lewis}, {Lidman}, {Macaulay},
  {M{\"o}ller}, {Nichol}, {Sako}, {Sommer}, {Swann}, {Tucker}, {Uddin},
  {Wiseman}, {Zhang}, {Abbott}, {Abdalla}, {Allam}, {Annis}, {Avila}, {Brooks},
  {Buckley-Geer}, {Burke}, {Carnero Rosell}, {Carrasco Kind}, {Carretero},
  {Castander}, {Cunha}, {Davis}, {De Vicente}, {Diehl}, {Doel}, {Eifler},
  {Flaugher}, {Fosalba}, {Frieman}, {Garc{\'{\i}}a-Bellido}, {Gruen},
  {Gruendl}, {Gutierrez}, {Hartley}, {Hollowood}, {Honscheid}, {James},
  {Jeltema}, {Kuropatkin}, {Li}, {Lima}, {Maia}, {Martini}, {Menanteau},
  {Ogando}, {Plazas}, {Roodman}, {Sanchez}, {Scarpine}, {Schindler}, {Smith},
  {Soares-Santos}, {Sobreira}, {Suchyta}, {Swanson}, {Tarle}, {Tucker}, \&
  {Walker}}]{Pursiainen18}
{Pursiainen}, M., {Childress}, M., {Smith}, M., {et~al.} 2018, MNRAS, 481, 894

\bibitem[{{Quataert} {et~al.}(2019){Quataert}, {Lecoanet}, \&
  {Coughlin}}]{Quataert+19}
{Quataert}, E., {Lecoanet}, D., \& {Coughlin}, E.~R. 2019, \mnras, 485, L83

\bibitem[{{Rest} {et~al.}(2018){Rest}, {Garnavich}, {Khatami}, {Kasen},
  {Tucker}, {Shaya}, {Olling}, {Mushotzky}, {Zenteno}, {Margheim},
  {Strampelli}, {James}, {Smith}, {F{\"o}rster}, \& {Villar}}]{Rest18}
{Rest}, A., {Garnavich}, P.~M., {Khatami}, D., {et~al.} 2018, Nature Astronomy,
  2, 307

\bibitem[{{Reynolds}(1999)}]{Reynolds99}
{Reynolds}, C.~S. 1999, in Astronomical Society of the Pacific Conference
  Series, Vol. 161, High Energy Processes in Accreting Black Holes, ed.
  J.~{Poutanen} \& R.~{Svensson}, 178

\bibitem[{{Risaliti} {et~al.}(2013){Risaliti}, {Harrison}, {Madsen}, {Walton},
  {Boggs}, {Christensen}, {Craig}, {Grefenstette}, {Hailey}, {Nardini},
  {Stern}, \& {Zhang}}]{Risaliti13}
{Risaliti}, G., {Harrison}, F.~A., {Madsen}, K.~K., {et~al.} 2013, \nat, 494,
  449

\bibitem[{{Rivera Sandoval} \& {Maccarone}(2018)}]{UVOTATel}
{Rivera Sandoval}, L.~E., \& {Maccarone}, T. 2018, The Astronomer's Telegram,
  11737, 1

\bibitem[{{Rivera Sandoval} {et~al.}(2018){Rivera Sandoval}, {Maccarone},
  {Corsi}, {Brown}, {Pooley}, \& {Wheeler}}]{Riv18}
{Rivera Sandoval}, L.~E., {Maccarone}, T.~J., {Corsi}, A., {et~al.} 2018,
  \mnras, 480, L146

\bibitem[{{Rosen} {et~al.}(2016){Rosen}, {Webb}, {Watson}, {Ballet}, {Barret},
  {Braito}, {Carrera}, {Ceballos}, {Coriat}, {Della Ceca}, {Denkinson},
  {Esquej}, {Farrell}, {Freyberg}, {Gris{\'e}}, {Guillout}, {Heil},
  {Koliopanos}, {Law-Green}, {Lamer}, {Lin}, {Martino}, {Michel}, {Motch},
  {Nebot Gomez-Moran}, {Page}, {Page}, {Page}, {Pakull}, {Pye}, {Read},
  {Rodriguez}, {Sakano}, {Saxton}, {Schwope}, {Scott}, {Sturm}, {Traulsen},
  {Yershov}, \& {Zolotukhin}}]{Rosen2016}
{Rosen}, S.~R., {Webb}, N.~A., {Watson}, M.~G., {et~al.} 2016, \aap, 590, A1

\bibitem[{{Sadowski} \& {Narayan}(2015)}]{Sadowski&Narayan15}
{Sadowski}, A., \& {Narayan}, R. 2015, \mnras, 453, 3213

\bibitem[{{Sadowski} \& {Narayan}(2016)}]{Sadowski&Narayan16}
---. 2016, \mnras, 456, 3929

\bibitem[{{Sagiv} {et~al.}(2014){Sagiv}, {Gal-Yam}, {Ofek}, {Waxman},
  {Aharonson}, {Kulkarni}, {Nakar}, {Maoz}, {Trakhtenbrot}, {Phinney}, {Topaz},
  {Beichman}, {Murthy}, \& {Worden}}]{Sagiv14}
{Sagiv}, I., {Gal-Yam}, A., {Ofek}, E.~O., {et~al.} 2014, \aj, 147, 79

\bibitem[{{Schr{\o}der} {et~al.}(2020){Schr{\o}der}, {MacLeod}, {Loeb},
  {Vigna-G{\'o}mez}, \& {Mandel}}]{Schroder+20}
{Schr{\o}der}, S.~L., {MacLeod}, M., {Loeb}, A., {Vigna-G{\'o}mez}, A., \&
  {Mandel}, I. 2020, \apj, 892, 13

\bibitem[{{Shakura} \& {Sunyaev}(1973)}]{Shakura&Sunyaev73}
{Shakura}, N.~I., \& {Sunyaev}, R.~A. 1973, \aap, 500, 33

\bibitem[{{Smartt} {et~al.}(2018){Smartt}, {\noopsort{a}}{Clark}, {Smith},
  {McBrien}, {Maguire}, {O'Neil}, {Fulton}, {Magee}, {Prentice}, {Colin},
  {Tonry}, {Denneau}, {Stalder}, {Heinze}, {Weiland}, {Flewelling}, \&
  {Rest}}]{Smartt+2018a}
{Smartt}, S.~J., {\noopsort{a}}{Clark}, P., {Smith}, K.~W., {et~al.} 2018, The
  Astronomer's Telegram, 11727

\bibitem[{{Soker} {et~al.}(2019){Soker}, {Grichener}, \& {Gilkis}}]{Soker+19}
{Soker}, N., {Grichener}, A., \& {Gilkis}, A. 2019, \mnras, 484, 4972

\bibitem[{{Stone} {et~al.}(2013){Stone}, {Sari}, \& {Loeb}}]{Stone+13}
{Stone}, N., {Sari}, R., \& {Loeb}, A. 2013, \mnras, 435, 1809

\bibitem[{{Sun} {et~al.}(2022){Sun}, {Maund}, {Crowther}, \& {Liu}}]{Sun22}
{Sun}, N.-C., {Maund}, J.~R., {Crowther}, P.~A., \& {Liu}, L.-D. 2022, \mnras,
  512, L66

\bibitem[{{Sun} {et~al.}(2023){Sun}, {Maund}, {Shao}, \& {Janiak}}]{Sun22b}
{Sun}, N.-C., {Maund}, J.~R., {Shao}, Y., \& {Janiak}, I.~A. 2023, \mnras, 519,
  3785

\bibitem[{{Tampo} {et~al.}(2020){Tampo}, {Tanaka}, {Maeda}, {Yasuda},
  {Tominaga}, {Jiang}, {Moriya}, {Morokuma}, {Suzuki}, {Takahashi}, {Kokubo},
  \& {Kawana}}]{Tampo+2020}
{Tampo}, Y., {Tanaka}, M., {Maeda}, K., {et~al.} 2020, \apj, 894, 27

\bibitem[{{Thomas} {et~al.}(2022){Thomas}, {Wheeler}, {Dwarkadas}, {Stockdale},
  {Vink{\'o}}, {Pooley}, {Xu}, {Zeimann}, \& {MacQueen}}]{Thomas22}
{Thomas}, B.~P., {Wheeler}, J.~C., {Dwarkadas}, V.~V., {et~al.} 2022, \apj,
  930, 57

\bibitem[{{Thomsen} {et~al.}(2019){Thomsen}, {Lixin Dai}, {Ramirez-Ruiz},
  {Kara}, \& {Reynolds}}]{Thomsen2019}
{Thomsen}, L.~L., {Lixin Dai}, J., {Ramirez-Ruiz}, E., {Kara}, E., \&
  {Reynolds}, C. 2019, \apjl, 884, L21

\bibitem[{{Traulsen} {et~al.}(2019){Traulsen}, {Schwope}, {Lamer}, {Ballet},
  {Carrera}, {Coriat}, {Freyberg}, {Michel}, {Motch}, {Rosen}, {Webb},
  {Ceballos}, {Koliopanos}, {Kurpas}, {Page}, \& {Watson}}]{Trau19}
{Traulsen}, I., {Schwope}, A.~D., {Lamer}, G., {et~al.} 2019, \aap, 624, A77

\bibitem[{{Traulsen} {et~al.}(2020){Traulsen}, {Schwope}, {Lamer}, {Ballet},
  {Carrera}, {Ceballos}, {Coriat}, {Freyberg}, {Koliopanos}, {Kurpas},
  {Michel}, {Motch}, {Page}, {Watson}, \& {Webb}}]{Trau20}
---. 2020, \aap, 641, A137

\bibitem[{{Tuna} \& {Metzger}(2023)}]{TunaMetzger23}
{Tuna}, S., \& {Metzger}, B.~D. 2023, arXiv e-prints, arXiv:2306.10111

\bibitem[{{van Velzen} {et~al.}(2019){van Velzen}, {Stone}, {Metzger},
  {Gezari}, {Brown}, \& {Fruchter}}]{vanVelzen19}
{van Velzen}, S., {Stone}, N.~C., {Metzger}, B.~D., {et~al.} 2019, \apj, 878,
  82

\bibitem[{{Virtanen} {et~al.}(2020){Virtanen}, {Gommers}, {Oliphant},
  {Haberland}, {Reddy}, {Cournapeau}, {Burovski}, {Peterson}, {Weckesser},
  {Bright}, {van der Walt}, {Brett}, {Wilson}, {Millman}, {Mayorov}, {Nelson},
  {Jones}, {Kern}, {Larson}, {Carey}, {Polat}, {Feng}, {Moore}, {VanderPlas},
  {Laxalde}, {Perktold}, {Cimrman}, {Henriksen}, {Quintero}, {Harris},
  {Archibald}, {Ribeiro}, {Pedregosa}, {van Mulbregt}, \& {SciPy 1. 0
  Contributors}}]{scipy}
{Virtanen}, P., {Gommers}, R., {Oliphant}, T.~E., {et~al.} 2020, Nature
  Methods, 17, 261

\bibitem[{{Vurm} \& {Metzger}(2021)}]{Vurm&Metzger21}
{Vurm}, I., \& {Metzger}, B.~D. 2021, \apj, 917, 77

\bibitem[{{Webb} {et~al.}(2020){Webb}, {Coriat}, {Traulsen}, {Ballet}, {Motch},
  {Carrera}, {Koliopanos}, {Authier}, {de la Calle}, {Ceballos}, {Colomo},
  {Chuard}, {Freyberg}, {Garcia}, {Kolehmainen}, {Lamer}, {Lin}, {Maggi},
  {Michel}, {Page}, {Page}, {Perea-Calderon}, {Pineau}, {Rodriguez}, {Rosen},
  {Santos Lleo}, {Saxton}, {Schwope}, {Tom{\'a}s}, {Watson}, \&
  {Zakardjian}}]{webb20}
{Webb}, N.~A., {Coriat}, M., {Traulsen}, I., {et~al.} 2020, \aap, 641, A136

\bibitem[{{Xiang} {et~al.}(2021){Xiang}, {Wang}, {Lin}, {Mo}, {Lin}, {Burke},
  {Hiramatsu}, {Hosseinzadeh}, {Howell}, {McCully}, {Valenti}, {Vink{\'o}},
  {Wheeler}, {Ehgamberdiev}, {Mirzaqulov}, {B{\'o}di}, {Bogn{\'a}r}, {Cseh},
  {Hanyecz}, {Ign{\'a}cz}, {Kalup}, {K{\"o}nyves-T{\'o}th}, {Kriskovics},
  {Ordasi}, {P{\'a}l}, {S{\'a}rneczky}, {Seli}, {Szak{\'a}ts}, {Arranz-Heras},
  {Benavides-Palencia}, {Cejudo-Mart{\'\i}nez}, {De la Fuente-Fern{\'a}ndez},
  {Escart{\'\i}n-P{\'e}rez}, {Garc{\'\i}a-De la Cuesta},
  {Gonz{\'a}lez-Carballo}, {Gonz{\'a}lez-Farf{\'a}n},
  {Lim{\'o}n-Mart{\'\i}nez}, {Mantero}, {Naves-Nogu{\'e}s}, {Morales-Aimar},
  {Ru{\'\i}z-Ru{\'\i}z}, {Sold{\'a}n-Alfaro}, {Valero-P{\'e}rez},
  {Violat-Bordonau}, {Zhang}, {Zhang}, {Li}, {Chen}, {Sai}, \&
  {Li}}]{Xiang21-18cow}
{Xiang}, D., {Wang}, X., {Lin}, W., {et~al.} 2021, \apj, 910, 42

\bibitem[{{Yao} {et~al.}(2022){Yao}, {Ho}, {Medvedev}, {Nayana}, {Perley},
  {Kulkarni}, {Chandra}, {Sazonov}, {Gilfanov}, {Khorunzhev}, {Khatami}, \&
  {Sunyaev}}]{Yao21}
{Yao}, Y., {Ho}, A. Y.~Q., {Medvedev}, P., {et~al.} 2022, \apj, 934, 104

\bibitem[{{Yuan} \& {Narayan}(2014)}]{Yuan&Narayan14}
{Yuan}, F., \& {Narayan}, R. 2014, \araa, 52, 529

\bibitem[{{Zhang} {et~al.}(2022){Zhang}, {Shu}, {Chen}, {Sun}, {Shen}, {Tao},
  {Chen}, {Jiang}, {Dou}, {Qin}, {Zhang}, {Zhang}, {Qu}, \&
  {Wang}}]{Zhang+2022}
{Zhang}, W., {Shu}, X., {Chen}, J.-H., {et~al.} 2022, Research in Astronomy and
  Astrophysics, 22, 125016

\end{thebibliography}



\end{document}